\documentclass[prx,aps,twocolumn,amsmath,amssymb]{revtex4-1}

\pdfoutput=1
\usepackage{hyperref}
\usepackage{graphicx}
\usepackage[usenames]{color}
\usepackage{bm}
\usepackage[final]{movie15}

\definecolor{blue}{rgb}{0,0,1}

\def\eqq#1{Eq.~(\ref{#1})}
\def\eq#1{(\ref{#1})}
\def\av#1{\langle #1 \rangle}

\def\f#1{Fig.~\ref{#1}}

\def\c#1{~\cite{#1}}
\def\cc#1{Ref.\c{#1}}

\def\s#1{Section~\ref{#1}}

\def\e{{\rm e}}

\def\beq{\begin{equation}}
\def\eeq{\end{equation}}
\def\bea{\begin{eqnarray}}
\def\eea{\end{eqnarray}}

\begin{document}

\title{Direct evaluation of dynamical large-deviation rate functions using a variational ansatz}
\author{Daniel Jacobson$^1$}
\email{{\tt Daniel.Jacobson@caltech.edu}}
\author{Stephen Whitelam$^2$}
\email{{\tt swhitelam@lbl.gov}}
\affiliation{$^1$Division of Chemistry and Chemical Engineering, California Institute of Technology, Pasadena, California 91125, USA\\
$^2$Molecular Foundry, Lawrence Berkeley National Laboratory, 1 Cyclotron Road, Berkeley, CA 94720, USA}
\begin{abstract}
We describe a simple form of importance sampling designed to bound and compute large-deviation rate functions for time-extensive dynamical observables in continuous-time Markov chains. We start with a model, defined by a set of rates, and a time-extensive dynamical observable. We construct a reference model, a variational ansatz for the behavior of the original model conditioned on atypical values of the observable. Direct simulation of the reference model provides an upper bound on the large-deviation rate function associated with the original model, an estimate of the tightness of the bound, and, if the ansatz is chosen well, the exact rate function. The exact rare behavior of the original model does not need to be known in advance. We use this method to calculate rate functions for currents and counting observables in a set of network- and lattice models taken from the literature. Straightforward ans\"{a}tze yield bounds that are tighter than bounds obtained from Level 2.5 of large deviations via approximations that involve uniform scalings of rates. We show how to correct these bounds in order to recover the rate functions exactly. Our approach is complementary to more specialized methods, and offers a physically transparent framework for approximating and calculating the likelihood of dynamical large deviations.
\end{abstract}

\maketitle

\section{Introduction} 

Dynamical systems, such as chemical networks\c{gillespie2007stochastic}, biochemical and molecular machines\c{mcgrath2017biochemical,seifert2012stochastic,brown2017allocating}, and models of driven\c{visco2006fluctuations,chou2011non,vaikuntanathan2014dynamic, harris2015fluctuations} and glassy\c{berthier2014nonequilibrium,garrahan2007dynamical,garrahan2009first} systems, exhibit fluctuations, departures from typical behavior\c{ritort2008nonequilibrium}. Fluctuations of time-extensive observables -- which can be work, entropy production\c{seifert2005entropy,speck2012large}, other currents\c{lecomte2010current,harris2015fluctuations,gingrich2017fundamental}, or dynamical activity\c{garrahan2009first,fodor2015activity} -- characterize the behavior of these systems, much as fluctuations of size-extensive quantities, such as energy or magnetization, characterize the static behavior of equilibrium systems\c{chandler1987introduction,binney1992theory}. The probability distributions that control dynamical fluctuations satisfy certain requirements, known as fluctuation relations\c{gallavotti1995dynamical,maes1999fluctuation,jarzynski1997nonequilibrium,kurchan1998fluctuation,lebowitz1999gallavotti,crooks1999entropy,seifert2005entropy,seifert2012stochastic,harris2007fluctuation,speck2012large}, which impose constraints on their symmetries.  The precise form of these distributions, however, must be obtained by explicit calculation. 

Here we focus on calculating probability distributions $\rho_T(A)$ for models with a discrete state space, for stochastic dynamical trajectories of elapsed time $T$ and time-extensive observables $A$. Time-extensive observables are those that can be built from a sum of values of individual pieces of a trajectory. For large values of $T$ these distributions often adopt the large-deviation form\c{gallavotti1995dynamical,maes1999fluctuation,jarzynski1997nonequilibrium,kurchan1998fluctuation,lebowitz1999gallavotti,crooks1999entropy,den2008large,touchette2009large}
\beq
\label{d1}
\rho_T(A) \approx \e^{-T J(a)},
\eeq 
in which $a = A/T$ is the time-intensive value of the observable. $J(a)$ is the large-deviation {\em rate function}, which quantifies the likelihood of observing particular values of the observable $a$~\footnote{More directly it quantifies the rate of decay of a fluctuation $a$, which depends both on the likelihood of $a$ {\em and} the basic timescale governing the establishment and decay of fluctuations. This latter piece plays a key role for certain models\c{spiliopoulos2013large,bouchet2016large,whitelam2018large}.}. The symbol $\approx$ denotes equality of the logarithms of both sides of \eq{d1}, to leading order in $T$, for all values of $a$. In the physics literature, most numerical methods for calculating $J(a)$ aim to first compute its Legendre transform, the scaled cumulant-generating function (SCGF) \c{giardina2006direct,touchette2009large,garrahan2009first,chetrite2015variational,jack2015effective,ray2018exact,nemoto2017finite,ferre2018adaptive}. It is possible to recover $J(a)$ from the SCGF if the former is convex\c{chetrite2013nonequilibrium}. A common way to calculate the SCGF is to use cloning methods\c{giardina2006direct,lecomte2007numerical}, which duplicate or eliminate trajectories according to their time-integrated weights. Often cloning is supplemented by other importance-sampling methods\c{jack2015effective,ray2018exact,nemoto2017finite,ray2018exact}, some of which make use of a modified dynamics in order to produce trajectories that more closely resemble the rare dynamics of the original model. 

Determining $J(a)$ {\em solely} by reweighting trajectories of a modified dynamics, without prior knowledge of the rare dynamics of the model of interest, is not widely done (see, however, Refs.\c{kundu2011application,klymko2017rare,whitelam2018sampling,whitelam2018multi}). Standard arguments suggest that determining the probability distribution of $a$ within one dynamics by reweighting against a second dynamics requires, in general, the evaluation of random quantities whose variance is exponential in the trajectory length\c{bucklew1990monte,glynn1989importance,sadowsky1990large,glasserman1997counterexamples} (see \s{efficiency}). Such observations are sometimes taken to mean that trajectory reweighting, without advance knowledge of the rare dynamics to be sampled, is little better than direct sampling using the original model\c{warren2018trajectory}. Here we argue that more optimism is warranted, and show that the conditions under which meaningful results can be extracted from trajectory reweighting are much less restrictive than has been recognized. Moreover, trajectory reweighting presents few technical complications beyond the requirement to simulate the original model with modified rates, and allows the reconstruction of $J(a)$ directly, without first calculating the SCGF.

To compute the large-deviation rate function $J(a)$ for a given model and dynamical observable $a$, we use a simple form of importance sampling\c{glynn1989importance,sadowsky1990large,bucklew1990monte,bucklew1990large,asmussen2007stochastic,juneja2006rare,bucklew2013introduction}. We begin with a modification of the model dynamics. This modified or {\em reference} model is a microscopic ansatz for the original model's behavior, conditioned on particular values of the observable $a$. The ansatz is characterized by a set of parameters whose values we determine variationally. In practical terms we simply guess a reference dynamics that is able to generate more or less of $a$ than the normal dynamics. Let the typical value of $a$ produced by original and reference models be $a_0$ and $\tilde{a}_0$, respectively~\footnote{The ``typical'' value of a time-integrated observable is the value to which the sample mean of an unbiased estimator of that quantity converges at long times, provided that a large-deviation principle exists and that the associated rate function has a unique zero\c{touchette2009large}.}. Reweighting trajectories of the reference model produces an upper bound on the rate function $J$ associated with the original model at the point $\tilde{a}_0$, an estimate of the tightness of the bound, and, if the ansatz is chosen well, the exact rate function (to within statistical error). That is, the reference dynamics is a true ansatz, a guess whose accuracy can be determined by subsequent calculation. Repeating the calculation for a set of reference models possessing a set of distinct values $\{\tilde{a}_0\}$ allows us to attempt reconstruction of $J(a)$ at the set of points $a \in \{\tilde{a}_0\}$. In this respect the procedure is similar to umbrella sampling of equilibrium systems. We show that the conditions under which the exact rate function can be recovered are less restrictive than usually assumed.

Any reference dynamics can be reweighted to produce {\em some} upper bound on $J(a)$, simply by making the desired value of $a$ typical\c{varadhan2010large}. Good choices of reference dynamics, leading to tight bounds, render the fluctuations of the reweighting factor or likelihood ratio (the ratio of path probability of new and old dynamics) small. We show here that relatively simple reference-model choices produce meaningful (i.e. tight) bounds, for a set of models and observables taken from the literature. We compare the bounds produced by our method with universal bounds on currents\c{pietzonka2016universal,gingrich2016dissipation} and non-decreasing counting variables\c{garrahan2017simple}. Those bounds can be obtained from Level 2.5 of large deviations\c{maes2008canonical,bertini2015large}, the exact rate function for the empirical flow (jumps between states) and measure (state occupation times), via a uniform rescaling of rates. That approach provides important physical insight into the quantities that constrain fluctuations of time-integrated observables, and also provides numerical bounds on rate functions. Our approach, which uses a microscopic ansatz within the exact path integral for the dynamics, produces tighter bounds, particularly far into the tails of rate functions. The extent to which bounds vary as we change the nature of the ansatz provides physical insight into how much certain types of microscopic processes contribute to the rare dynamics of a model. Microscopic ans\"atze, even relatively simple ones, are capable of capturing a wide range of behavior, including regimes of anomalous fluctuations in which the usual central-limit theorem breaks down\c{klymko2017rare}. Computing a correction to these bounds, by measuring fluctuations of the likelihood ratio, allows the recovery of the exact rate function. Importantly, fluctuations of the likelihood ratio do not need to be zero for $J(a)$ to be calculated

The approach described here is variational, in the sense that we vary the parameters of the reference model in order to identify the dynamics that best approximates the rare dynamics of interest. Variational principles underpin the study of large deviations, embodied by the notion that ``any large deviation is done in the least unlikely of all the unlikely ways''\c{den2008large}. Variational ideas are central to different representations of rare processes -- see e.g. Section 5 of \cc{chetrite2015variational} -- and have been widely used in analytic and numerical work\c{PhysRevE.94.032101,jack2013large,ray2017importance, ray2018exact,ferre2018adaptive}. The aim of this paper is to present a simple, physically-motivated approach to bounding and calculating rate functions using a variational principle enacted by (only) direct simulations, and to present a set of convergence criteria, adapted from \cc{rohwer2015convergence}, that reveal when bounds can be corrected to produce the exact rate function. We have provided a GitHub script\c{jacobson2019vard} that computes the correction term automatically, this being the most involved step of the calculation. These results extend our previous work\c{klymko2017rare,whitelam2018sampling,whitelam2018multi} by a) showing how different forms of physically-motivated reference dynamics can be used to treat different models, and b) by providing a set of criteria that identify when the exact rate function can be recovered. One point we emphasize is that considerable progress can be made using physical intuition and basic knowledge of the properties of a model, without the application of other forms of importance sampling (such as cloning or transition-path sampling). Our method requires only continuous-time Monte Carlo simulation, and so can be applied to any set of circumstances in which that method can be used, including to models with unbounded state spaces\c{klymko2017rare}. In addition, it can be used to reconstruct families of large-deviation rate functions from a single set of simulations, using the principle that the dynamics of one model can be reweighted to examine the dynamics of many others\c{whitelam2018multi}. Reference models represent a form of importance sampling similar in spirit but different in detail to the umbrella potentials used in equilibrium sampling\c{torrie1977nonphysical,tenWolde_1998}.\\
~\\
\indent In \s{desc} we describe our approach, which we refer to as VARD (for Variational Ansatz for Rare Dynamics). In general terms there are many forms of VARD that have been used in the literature (see above); we use the term to convey the specific notion of doing (only) direct simulations of a family of modified models. In \s{app} we apply the method to four models taken from the literature. We have chosen models from the literature that display a variety of interesting behavior: two lattice models (the asymmetric simple exclusion process\c{derrida1998exactly,chou2011non} and the Fredrickson-Andersen model\c{fredrickson1984kinetic}) and two network models, and we sample both currents and non-decreasing counting variables to show that the method works the same way for each. In the cases described in \s{four_state}--\s{sec_fa}, relatively simple choices of reference model allow the computation of the exact rate function, and we gain physical insight into the nature of the dynamics that contributes to particular pieces of $J(a)$. We also show, in \s{large}, that bounds that are descriptive in small systems remain so in larger systems. We conclude in \s{conc}.

\section{A variational ansatz for rare dynamics}
\label{desc}

\subsection{Continuous-time Markov chains and large deviations}

Consider a continuous-time dynamics\c{binder1986introduction} on a set of discrete states, defined by the master equation
\beq
\label{me}
\partial_t P_x(t) = \sum_{y\neq x} W_{yx} P_y(t)- R_x P_x(t).
\eeq
Here $P_x(t)$ is the probability that a system resides in (micro)state $x$ at time $t$. $W_{xy}$ is the rate for passing from state $x$ to state $y$, and $R_x = \sum_{y \neq x} W_{xy}$ is the escape rate from $x$ (Table 1 provides a reference for some of the more frequently-used symbols in this paper). The standard algorithm for simulating the dynamics \eq{me} is as follows\c{gillespie1977exact}. From state $x$, choose a destination state $y$ with probability
\beq
\label{one}
p_{xy} = \frac{W_{xy}}{R_x}.
\eeq
The time increment $\Delta t$ required to make this move is a random number drawn from the exponential distribution with mean $1/R_x$,
\beq
\label{two}
p_{x}(\Delta t) = R_x \e^{-R_x \Delta t}.
\eeq
Given an initial state $x_0$, the dynamics defined by \eq{one} and \eq{two} generates a trajectory $\omega=x_0 \to x_1 \to \dots \to x_{K(\omega)}$, which consists of a sequence of $K(\omega)$ jumps $x_k \to x_{k+1}$ and associated jump times $\Delta t_k$. In this paper we are concerned with calculating
\beq
\label{av}
\rho_T(A)=\sum_{\omega} p(\omega) \delta{(T(\omega)-T)} \delta{(A(\omega)-A)},
\eeq
the probability distribution, taken over all trajectories of elapsed time $T$, of a time-extensive dynamical observable 
\beq
\label{ay}
A(\omega) = \sum_{k=0}^{K(\omega)-1} \alpha_{x_k x_{k+1}}.
\eeq
Here $\alpha_{xy}$ is the change of the observable $A$ upon moving from $x$ to $y$, and $A(\omega)$ is the sum of these quantities over a single trajectory $\omega$. We define $a(\omega) \equiv A(\omega) / T(\omega)$ as the time-intensive version of $A$. $T(\omega)$ is the elapsed time of trajectory $\omega$, and is equal to $T$ when $T_{K(\omega)} \leq T < T_{K(\omega)+1}$, where $ T_{K(\omega)}=\sum_{k=0}^{K(\omega)-1} \Delta t_k$. The symbol $p(\omega)$ is the probability of a trajectory $\omega$. Given an initial state, this term is equal to a product of factors \eq{one} and \eq{two} for all jumps of the trajectory, multiplied by the probability of not exiting state $x_{K(\omega)}$ between times $T_{K(\omega)}$ and $T$.

\begin{table}[h]
\caption{Glossary of frequently-used symbols}
\begin{center}
\begin{tabular}{cl}
$A=aT$ & path-extensive order parameter\\
$T$ & elapsed time of a trajectory\\
$K$ & number of jumps of a trajectory \\
$\alpha_{xy}$ & change of $A$ upon making the jump $x \to y$ \\
$W_{xy}$ & ``original'' model rates for the jump $x\to y$\\
$R_x$ & ``original'' model escape rate $\sum_{y\neq x}W_{xy}$ \\
$a_0$ & typical value of $a$ under the original dynamics\\
$J(a)$ & rate function for $a$ under the original dynamics\\\
$J_0(a)$ & upper bound on $J(a)$\\
$J_1(a)$ & correction to the bound: $J(a)=J_0(a) + J_1(a)$\\
$\tilde{W}_{xy}$ & reference-model rates for jumps $x\to y$\\
$s$ & one parameter of $\tilde{W}_{xy}$: see \eq{rt} \\
$\beta_{xy}$ & remaining parameters of $\tilde{W}_{xy}$: see \eq{rt} \\
$\tilde{R}_x$ & reference-model escape rate $\sum_{y\neq x}\tilde{W}_{xy}$ \\
$\lambda$ & clock bias: see \eq{four}\\
$\tilde{a}_0$ & typical value of $a$ under the reference dynamics\\
$J_0[\{x\}]$ & $J_0(a)$ determined from a scan of the \\ 
\, & reference-model parameter set $\{x\}$ 
\end{tabular}
\end{center}
\label{default}
\end{table}

In \eq{av}, the delta functions denote the conditions of fixed $A$ and fixed $T$ that we wish to impose on the trajectory ensemble. This conditioning defines the {\em microcanonical path ensemble}\c{chetrite2015variational}, of which \eq{av} is the normalization constant.

\begin{figure*}[] 
   \centering
  \includegraphics[width=\linewidth]{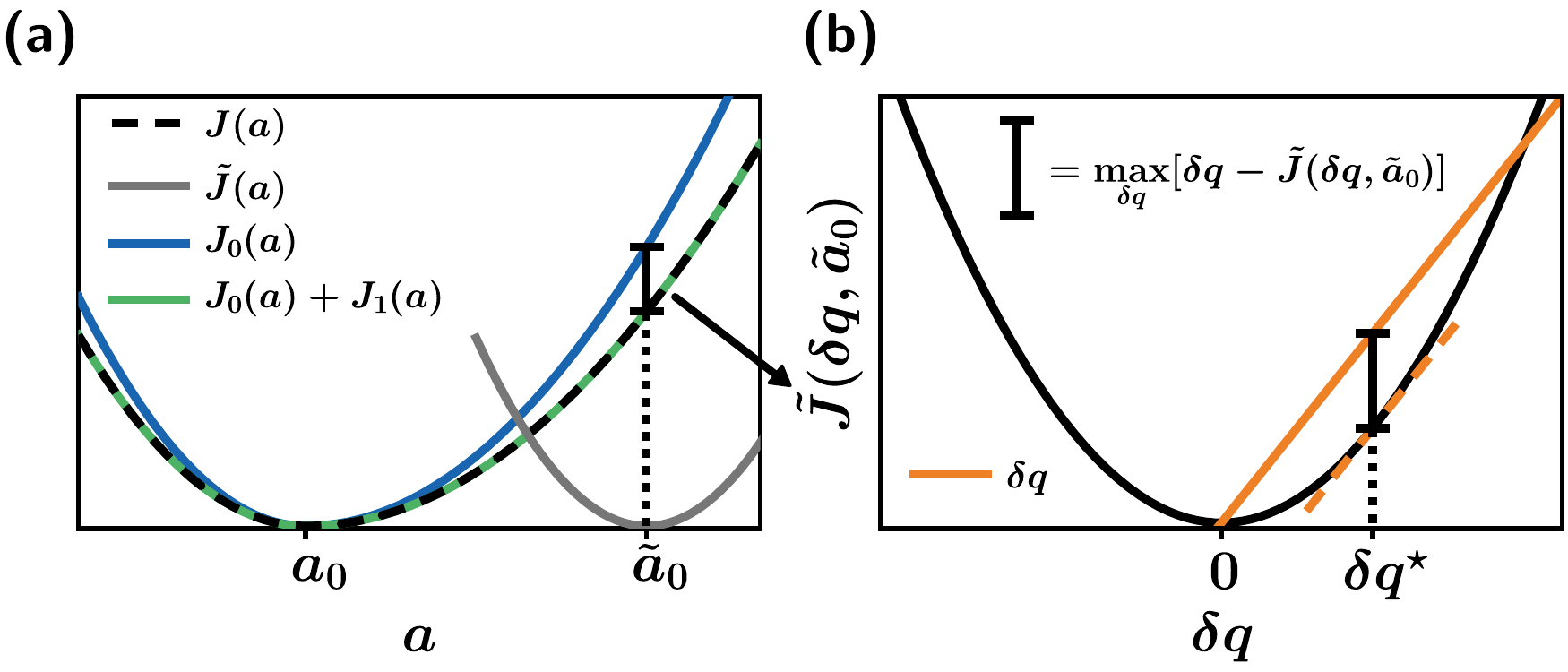} 
  \caption{Large deviations from a variational ansatz for rare dynamics (VARD). (a) We aim to bound or calculate the large-deviation rate function $J(a)$ (black dashed line) for a given model and observable $a$, under the continuous-time dynamics \eq{one} and \eq{two}. We introduce a reference model \eq{three} and \eq{four}, a variational ansatz for the rare dynamics of the original model conditioned upon $a$. The reference model has rate function $\tilde{J}(a)$ (gray line; we do not aim to compute this function). The typical value of $a$ generated by the reference model is $\tilde{a}_0$; this is potentially far from the typical value $a_0$ generated by the original model. Evaluation of \eq{j0} from the sample mean of a single trajectory of the reference model produces one point $J_0(\tilde{a}_0)$ on the blue line, an upper bound on $J(a)$. (b) If we can evaluate the auxiliary rate function $\tilde{J}(\delta q, {\tilde{a}_0})$ at the point $\delta q^\star$ at which its gradient is unity, then we can calculate the correction term \eq{j1} and determine one point on the green curve in panel (a), the exact rate function of the original model. If not, then we gain information about the quality of the bound $J_0(\tilde{a}_0)$. Variation of the parameters of the reference model allows reconstruction of the entire blue and (potentially) green curves in panel (a). VARD thus reduces a single nonlocal problem, the computation of $J(a)$ arbitrarily far from $a_0$, to a series of local problems, each requiring the evaluation of an auxiliary rate function $\tilde{J}(\delta q, {\tilde{a}_0})$ at the point $\delta q^\star$ at which its gradient is unity. As we show in this paper, this procedure can be carried out for models commonly found in the literature using relatively simple choices of reference model.}
   \label{fig0}
\end{figure*}

We focus on models for which, for large values of $T$, the probability distribution \eq{av} adopts the large-deviation form \eq{d1}. Our aim is to calculate the rate function $J(a)$ (sometimes the notation $I(a)$ or $\varphi(a)$ is used to denote a rate function\c{touchette2009large,garrahan2017simple}). This function quantifies the rate of decay of atypical values of $a$. For many models $J(a)$ has a unique minimum at a point $a=a_0$, where $J(a_0)=0$. This point defines the typical value of $a$: the distribution $\rho_T(A)$ concentrates on $a_0$ in the long-time limit, an expression of the law of large numbers\c{den2008large,touchette2009large}. In general, a rate function can have more than one point at which it is zero, defining multiple typical values of an observable\c{dinwoodie1992large,touchette2009large,klymko2017rare}. Often, the rate function is quadratic in the neighborhood of its minimum, an expression of the central-limit theorem\c{den2008large,touchette2009large}. Exceptions to this norm occur at phase transitions, where the usual central-limit theorem does not hold\c{ellis1985large}. Far from their minima, rate functions display a variety of behaviors\c{touchette2009large}. However, direct simulation of the dynamics \eq{one} and \eq{two} leads to poor sampling of $J(a)$ anywhere other than in the neighborhood $a \approx a_0$.

\subsection{Quantifying rare events}

To remedy the sampling problem for values of the observable $a$ far from $a_0$, we introduce a {\em reference model}. We wish to reweight the trajectories of the reference model in order to approximate or calculate $J(a)$ for values of $a$ potentially far from $a_0$. The reference model must satisfy certain requirements. It needs to be able to generate all trajectories possible in the original model, but no trajectories not possible in the original model (otherwise the reweighting factor, discussed below, can be infinity or zero). We want the reference model to be able to generate trajectories possessing values of $a$ that are rarely generated by the original model, which is relatively easy to arrange, but we also need to be able to recover from reference-model trajectories the probability with which such trajectories {\em would} have been generated by the original model. This second requirement is harder to arrange, but not prohibitively so. As we show, if the reference model generates trajectories possessing values of $a$ in a manner completely unlike the original model, then we have to do prohibitively heavy sampling of reference-model trajectories in order to calculate $J(a)$. If, however, the reference model generates trajectories possessing values of $a$ in a manner similar to the original model, then $J(a)$ can be reconstructed from trajectories of the reference model with relatively little effort. Importantly, the method tells us when this is so: we do not need to know in advance the precise nature of the rare dynamics of the original model in order to recover $J(a)$.

For a trajectory $\omega$ of the reference model we want to be able to influence how much of the dynamical observable is produced per jump, $A(\omega)/K(\omega)$, and the number of jumps per unit time, $K(\omega)/T$. To control the former we use a reference-model dynamics that selects destination states with probability 
\beq
\label{three}
\tilde{p}_{xy} = \frac{\tilde{W}_{xy}}{\tilde{R}_x},
\eeq
where $\tilde{W}_{xy}$ is an effective rate, and $\tilde{R}_x = \sum_{y \neq x} \tilde{W}_{xy}$. (The true rates of the reference model are, from \eq{three} and \eq{four}, $\tilde{W}_{xy} (R_x+\lambda)/\tilde{R}_x$.) Here we use the parameterization 
\beq
\label{rt}
\tilde{W}_{xy}= \e^{-s \alpha_{xy}- \beta_{xy}} W_{xy},
\eeq 
which is a modification of \eq{one}. The factor $\e^{-s \alpha_{xy}}$ is chosen in order to guide the jump destination according to the change of the observable $\alpha_{xy}$ weighted by a parameter $s$. In general such a bias is not sufficient to control $A(\omega)/K(\omega)$, and so we also consider an additional arbitrary bias, $\beta_{xy}$. For the models considered here a simple and physically-motivated guess for what $\beta_{xy}$ should be is sufficient to produce a good reference model. We shall return to this point.

To control the number of jumps per unit time, $K(\omega)/T$, we draw times between jumps of the reference model from the distribution
\beq
\label{four}
\tilde{p}_{x}(\Delta t) = (R_x+\lambda) \e^{-(R_x+\lambda) \Delta t},
\eeq
where $\lambda>-\min_x R_x$ serves to make the jump time from a given state unusually large or small by the reckoning of the original model. This ``clock trick'' provides a simple way of sampling jump times without having those times appear explicitly in the reweighting factor\c{whitelam2018multi}. (The parameter $\beta_{xy}$ can also affect jump times indirectly, if, for example, it is chosen to be proportional to $R_y$, the escape rate from the destination state.) 

Next observe that the path weight $p(\omega)$ in \eq{av} can be written $\tilde{p}(\omega) \phi(\omega)$, where $\phi(\omega)=p(\omega)/\tilde{p}(\omega)$ is the reweighting factor, the ratio of weights of a trajectory $\omega$ in the original and reference models. $\phi(\omega)$ is also known as the likelihood ratio or the Radon-Nikodym derivative\c{chetrite2015variational,bucklew2013introduction}. For a jump $x \to y$ in time $\Delta t$, the reweighting factor is the product of \eq{one} and \eq{two}, divided by the product of \eq{three} and \eq{four}; for the entire trajectory $\omega$ we have
\beq
\label{phi}
\phi(\omega) = \e^{s A(\omega)+\lambda T(\omega)+ T q(\omega)},
\eeq
where 
\beq
\label{kew}
q(\omega) = \frac{1}{T} \sum_{k=0}^{K(\omega)-1} \left(\beta_{x_k x_{k+1}} + \ln \frac{\tilde{R}_{x_k}}{R_{x_k}+\lambda} \right)
\eeq
is the piece of $\phi$ that is not fixed by the delta-function constraints in \eq{av}. (The time-dependent piece of \eq{phi} produced by $K(\omega)$ jumps is $\e^{\lambda T_{K(\omega)}}$; the contribution from the final entry in the path weight, the probability of not jumping between time $T_{K(\omega)}$ and $T(\omega)$, leads to the factor of $\e^{\lambda T (\omega)}$ shown in \eq{phi}.)

We can then write \eq{av} in the form
\beq
\label{avp}
\frac{\rho_T(A)}{\tilde{\rho}_T(A)}=\e^{s A+\lambda T} \frac{\sum_{\omega} \tilde{p}(\omega) \e^{T q(\omega)} \delta{(T)} \delta{(A)}}{\sum_{\omega} \tilde{p}(\omega) \delta{(T)} \delta{(A)}},
\eeq
where $\tilde{\rho}_T(A)\approx \e^{-T \tilde{J}(a)}$ is the analog of \eq{av} for the reference model, and we have used the shorthand $\delta{(X)} \equiv \delta{(X(\omega)-X)}$. Replacing the sums over trajectories with an integral over trajectory weights gives
\beq
\label{av2}
\frac{\rho_T(A)}{\tilde{\rho}_T(A)}=\e^{s A+\lambda T} \int {\rm d}q \, \tilde{p}_T(q|a) \e^{T q},
\eeq
where $\tilde{p}_T(q|a)$ is the normalized probability distribution of $q(\omega)$ for trajectories of the reference model that have specified values of $a$ and $T$. For large $T$ we assume that this distribution will obey a large-deviation principle of its own. If so we can write, using the rules of conditional probability,
\beq
\label{qdist}
\tilde{p}_T(q|a) \approx  \e^{-T \tilde{J}(q|a)} = \e^{-T \left[\tilde{J}(q, a) - \tilde{J}(a)\right]},
\eeq
where $\tilde{J}(q, a)$ is the joint rate function for $q$ and $a$ within the reference model.

We next take the large-$T$ limit in \eq{av2}, replace all probability distributions with their large-deviation forms, and set $a=\tilde{a}_0$, the value typical of the reference model (such that $\tilde{J}(\tilde{a}_0)=0$). The result, upon taking logarithms, is
\beq
\label{int1}
J(\tilde{a}_0) =-s \tilde{a}_0 - \lambda - \lim_{T \to \infty} T^{-1} \ln \int {\rm d}q \,  \e^{T [q - \tilde{J}(q, \tilde{a}_{0})]}.
\eeq
Finally, we introduce $\delta q \equiv q - \tilde{q}_0$, where $\tilde{q}_{0}$ is the value of $q$ typical of the reference model. This value can be computed from a single reference-model trajectory (for a given set of parameters $s,\lambda, \beta_{xy}$). Extracting $\e^{T\tilde{q}_0}$ from the exponential in \eq{int1} and evaluating the integral using the saddle-point method yields
\beq
\label{rf}
J(\tilde{a}_0) = J_0(\tilde{a}_0)+J_1(\tilde{a}_0),
\eeq
where 
\beq
\label{j0}
J_0(\tilde{a}_0)=-s \tilde{a}_0 - \lambda -\tilde{q}_0
\eeq
and
\bea
\label{j1}
J_1(\tilde{a}_0)=-\max_{\delta q}[\delta q - \tilde{J}(\delta q, \tilde{a}_0)].
\eea
Eqs.\eq{rf}--\eq{j1} provide an exact representation of the rate function, if the probability distributions \eq{av} and \eq{qdist} adopt large-deviation forms~\footnote{In an abuse of notation we have, for brevity, changed from writing the joint rate function in the form $\tilde{J}(q,a)$ in \eq{int1} to $\tilde{J}(\delta q,a)$ in \eq{j1} and subsequently.}. \f{fig0} illustrates the relationship between Equations~\eq{rf},~\eq{j0}, and~\eq{j1}, which are central to the sampling scheme discussed in this paper.

\subsection{We can compute $J(a)$ as the sum of a bound and a correction}

The piece $J_0(\tilde{a}_0) \geq J(\tilde{a}_0)$, \eqq{j0}, is an upper bound on the rate function at the point $a=\tilde{a}_0$, by Jensen's inequality applied to \eq{av2}, and can be obtained from the sample mean of single trajectory of the reference model. It is always possible to calculate this term. If $\tilde{J}(a)$ is locally quadratic about $\tilde{a}_0$, meaning that the usual central-limit theorem holds~\footnote{Note that the reference model can be well behaved in this manner even when the {\em original} model exhibits anomalous fluctuations such that $J(a)$ is not quadratic about its minimum\c{klymko2017rare}.}, then errors in the computation of $\tilde{a}_0$ go as $ \sqrt{\av{(a-\tilde{a}_0)^2}} \sim T^{-1/2}$. The same is true for the computation of $\tilde{q}_0$. Thus statistical errors in the computation of the bound can be made negligible simply by computing \eq{j0} for a sufficiently long trajectory.
 
The term $J_1(\tilde{a}_0)$, \eqq{j1}, is a correction to the bound, and can be calculated by sampling values of $q$, \eqq{kew}, of trajectories of the reference model that have $a=\tilde{a}_0$. It is possible to calculate this term if the reference model is chosen well, but not if it is chosen badly.

The two terms in \eq{j1} describe a competition between the logarithmic weight $\delta q$ associated with reference-model trajectories that have atypical values of $q$, and the logarithmic probability $\tilde{J}(\delta q, \tilde{a}_0)$ of observing such trajectories. When $\tilde{J}(\delta q, \tilde{a}_0)$ is differentiable, \eq{j1} can be written
\beq
\label{jstar}
J_1(\tilde{a}_0)=-\delta q^\star +\tilde{J}(\delta q^\star, \tilde{a}_0),
\eeq
where 
\beq
\label{dJddq}
\partial_{\delta q} \tilde{J}(\delta q, \tilde{a}_0)|_{\delta q=\delta q^\star}=1.
\eeq
Thus we need to measure the value of $\tilde{J}(\delta q, \tilde{a}_0)$ at the point $\delta q^\star$ at which its gradient is unity, which will be a unique point when $\tilde{J}(\delta q, \tilde{a}_0)$ is convex. The sampling problem is now {\em localized}: instead of sampling $J(a)$ arbitrarly far from $a_0$ (using the original model), we need only sample a specific piece $\delta q^\star$ of an auxiliary rate function, $\tilde{J}(\delta q, \tilde{a}_0)$ (using the reference model). This fact, summarized in \f{fig0}, shows why the present scheme has the potential to be much more efficient than unbiased simulation, if the reference model is chosen well.

This sampling problem is still formidable in general. If the reference model is chosen badly, meaning that its typical trajectories have very different character to trajectories of the original model that have $a=\tilde{a}_0$, then the bound will be slack, meaning that $J_0(\tilde{a}_0) \gg J(\tilde{a}_0)$, and so $J_1(\tilde{a}_0)$ will be large. In this case $\tilde{J}(\delta q, \tilde{a}_0)$ will be broad around its minimum $\delta q =0$ (the variance of $\delta q$ will be large) and unreasonably heavy sampling using the reference model will be required to determine the point $\delta q^\star$ (because this corresponds to a rare event within the reference model).

However, for good choices of the reference model the opposite situation arises: the bound will be tight, meaning that $J_0(\tilde{a}_0) \approx J(\tilde{a}_0)$, and so $J_1(\tilde{a}_0)$ will be small. In this case the latter can be evaluated with reasonable numerical effort (in the examples that follow we can gather the required statistics of $q$ by sub-sampling a single trajectory of the reference model.)  If we can reconstruct $\tilde{J}(\delta q^\star, \tilde{a}_0)$ then we can calculate $J_1(\tilde{a}_0)$ and we have obtained the exact rate function. 

\subsection{Computing the correction} 
\label{corr}

Given a model and an observable $a$, we construct a reference model \eq{three} and \eq{four} so as to approximate or calculate $J(a)$. In \s{app} we provide a set of worked examples of this procedure. In general terms we simply guess which rates of the original model can be modified so as to produce more or less of $a$ than usual, and introduce a parameter ($s$, $\lambda$, or $\beta_{xy}$) able to control the rate in question. We do not know in advance which combination of these modified rates best approximates the way in which the original model produces rare values of $a$, but by running short trajectories of the reference model for different values of its parameters we can identify how this is done within the space of possibilities defined by the reference model. From the sample mean of each reference-model trajectory we obtain the values $\tilde{a}_0$ and $\tilde{q}_0$; plotting $\tilde{a}_0$ against $-s \tilde{a}_0 -\lambda -\tilde{q}_0$ for a range of values of reference-model parameters, and identifying the lower envelope of these points (conveniently calculated using a union of convex hull constructions), gives the bound $J_0(a)$ associated with that choice of reference model.

This bound is the starting point for our attempt to calculate the correction $J_1(a)$. The correction term can be interpreted as a measure of how close the typical dynamics of the reference model is to the desired rare dynamics of the original model. If $J_1(a)=0$ then typical trajectories of the reference model correspond exactly to trajectories of the original model conditioned on the relevant value $a$ of the order parameter. If $J_1$ is small then (slightly) atypical versions of reference-model trajectories correspond to the desired rare dynamics; and if $J_1$ is large (or cannot be calculated) then it is the very rare trajectories of the reference model that correspond to the desired rare dynamics of the original model.

In previous versions of our sampling method\c{klymko2017rare,whitelam2018sampling,whitelam2018multi} we used a cumulant expansion to evaluate the integral in \eq{int1}, giving, in place of \eq{j1},
\beq
\label{cumu}
J_1^{\rm approx}(\tilde{a}_0)=\frac{T}{2} \sigma_{\rm ref}^2+\frac{T^2}{6} \kappa_{\rm ref}+\cdots.
\eeq
Here $ \sigma_{\rm ref}^2 \propto 1/T$ is the variance of $\delta q$ over typical trajectories of the reference model (those having $a = \tilde{a}_0$), i.e. $\sigma_{\rm ref}^2 = \av{(\delta q)^2}_{\rm ref}$, and $\kappa_{\rm ref} = \av{(\delta q)^3}_{\rm ref} \propto 1/T^2$. \eqq{cumu} can give accurate results for the rate-function correction\c{klymko2017rare,whitelam2018sampling,whitelam2018multi}, but does not provide a self-consistent way of determining {\em when} the correction is accurate. At best we can determine that the first omitted term in the expansion \eq{cumu} is small, but this does not provide a proof of convergence.

In this paper we present an alternative way to calculate the correction term \eq{j1}, which builds upon methods designed to compute rate functions (or their SCGF Legendre duals) empirically\c{duffy2005large,rohwer2015convergence}. This process is more involved than the computations required to evaluate \eq{cumu}, but has the advantage of providing a set of clear convergence criteria and statistical error bars. This information reveals when we have converged \eq{j1}, and so have the exact rate function, and when we do not, thus turning the reference-model guess into a true ansatz. In the remainder of this section we describe the method we use to compute \eq{j1}. We have provided a GitHub script\c{jacobson2019vard} for calculating the correction automatically.

To obtain the correction we first assume that $\tilde{J}(\delta q, a)$ is differentiable, and so work with \eq{jstar} instead of \eq{j1}.
We then introduce the two-dimensional scaled cumulant-generating function (SCGF),
\beq
\label{2D_scgf}
\tilde{\theta}(k_{\delta q}, k_{a}) \equiv \lim_{T \to \infty} T^{-1} \ln \langle{\e^{T( k_{\delta q} \delta q + k_{a} a)}\rangle}_{\rm ref},
\eeq
where the angle brackets denote a trajectory ensemble average within the reference model. The SCGF \eq{2D_scgf} is related to $\tilde{J}(\delta q, a)$ through the double Legendre transform
\beq
\label{lt}
\tilde{J}(\delta q, a) = k_{\delta q} \delta q + k_{a} a - \tilde{\theta}(k_{\delta q}, k_a)
\eeq
where $k_{\delta q}$ and $k_{a}$ are conjugate to $\delta q$ and $a$ respectively. As a result, if we want to calculate $\tilde{J}(\delta q, a)$ at a single point, we can calculate the right-hand side of \eq{lt}. Doing so is much more efficient than attempting to reconstruct $\tilde{J}(\delta q, a)$ directly, for the reasons given in \s{efficiency}.

The formula for the correction \eq{jstar} depends on $\tilde{J}(\delta q^{\star}, \tilde{a}_{0})$, which we can get from \eq{lt}:
\beq
\label{lt_star}
\tilde{J}(\delta q^{\star}, \tilde{a}_{0}) = k_{\delta q^{\star}} \delta q^{\star} + k_{\tilde{a}_{0}} \tilde{a}_{0} - \tilde{\theta}(k_{\delta q^{\star}}, k_{\tilde{a}_0}).
\eeq
We can simplify this relationship by combining the implicit definition of $k_{\delta q}$ in the Legendre transform with \eq{dJddq} to get 
\beq
\label{def_kdqstar}
k_{\delta q^\star} = \partial_{\delta q} \tilde{J}(\delta q, \tilde{a}_0)|_{\delta q=\delta q^\star} = 1.
\eeq
Inserting \eq{def_kdqstar} into \eq{lt_star} yields
\beq
\label{eval_lt}
\tilde{J}(\delta q^\star, \tilde{a}_{0}) = \delta q^{\star} + k_{\tilde{a}_{0}} \tilde{a}_{0} - \tilde{\theta}(1, k_{\tilde{a}_0}).
\eeq
The quantity $\tilde{a}_{0}$ is the typical value of the observable in the reference model, and can be obtained from a single reference-model trajectory. The three other unknown quantities on the right-hand side of \eq{eval_lt} that are needed for the correction are $\delta q^{\star}$, $k_{\tilde{a}_0}$, and $\tilde{\theta}(1, k_{\tilde{a}_0})$.

To calculate these quantities we have to compute the value of the two-dimensional SCGF \eq{2D_scgf} at various points $(k_{\delta q}, k_{a})$. We  can do this using a simple extension of existing techniques developed to sample points on 1D SCGFs\c{duffy2005large,rohwer2015convergence}. Following \cc{rohwer2015convergence} we generate a single long trajectory of the reference model and sub-sample it into $M$ approximately independent blocks $\omega_{i}$ of length $T(\omega_{i}) = B$. Within each block we record the sample mean of $\delta q$ and $a$, which we write as $\delta q_i$ and $a_i$. $\tilde{\theta}(k_{\delta q}, k_{a})$ can be calculated from this data set using the estimator
\beq
\label{estimator}
\hat{\tilde{\theta}}(k_{\delta q}, k_{a}) = \frac{1}{B} \ln \left( \frac{1}{M} \sum_{i = 1}^{M} \e^{B( k_{\delta q} \delta q_{i} + k_{a} a_{i})} \right).
\eeq
\eqq{estimator} is guaranteed to converge to the exact value of the SCGF, $\tilde{\theta}(k_{\delta q}, k_{a})$, in the limit $M \to \infty$ and $B \to \infty$. The convergence properties of this estimator for finite $M$ and $B$ will be addressed in the next section, \ref{convergence}. For now we assume that we can obtain convergence as needed. Finally, note that by changing the values of $k_{\delta q}$ and $k_a$ in \eq{estimator} a single data set consisting of values of $a_i$ and $\delta q_i$, generated from a single long trajectory, can be used to recover many points $(k_{\delta q}, k_{a})$ on $\tilde{\theta}(k_{\delta q}, k_{a})$.

We now turn to the calculation of the three unknowns in \eq{eval_lt}, $\delta q^{\star}$, $k_{\tilde{a}_0}$ and $\tilde{\theta}(1, k_{\tilde{a}_0})$. First we use the relation
\beq
\label{calc_kta0}
\tilde{a}_0 = \partial_{k_a} \tilde{\theta}(k_{\delta q^\star} = 1, k_a) |_{k_a = k_{\tilde{a}_0}}
\eeq
to find $k_{\tilde{a}_0}$. We do so by calculating the SCGF, $\tilde{\theta}(k_{\delta q}, k_a)$, along a 1D slice through its 2D domain using the estimator \eq{estimator}. This slice is defined by fixing $k_{\delta q} = k_{\delta q^\star} = 1$ and varying $k_a$. We then use the method of finite differences to get $\partial_{k_a} \tilde{\theta}(k_{\delta q^\star} = 1, k_a)$ at each point $k_a$, and find the point that fulfills \eq{calc_kta0}. Once we know the value of $k_{\tilde{a}_0}$ we can calculate $\tilde{\theta}(1, k_{\tilde{a}_0})$, again using \eq{estimator}. Finally we can compute $\delta q^\star$  using the analog of \eq{calc_kta0} for $\delta q$,
\beq
\label{calc_dq_star}
\delta q^\star = \partial_{k_{\delta q}} \tilde{\theta}(k_{\delta q}, k_{\tilde{a}_0}) |_{k_{\delta q} = 1}.
\eeq
Inserting $\delta q^{\star}$, $k_{\tilde{a}_0}$, $\tilde{\theta}(1, k_{\tilde{a}_0})$ and $\tilde{a}_0$ into \eq{eval_lt} yields $\tilde{J}(\delta q^{\star}, \tilde{a}_0)$. The correction to the bound, $J_{1}(\tilde{a}_0)$, then follows from \eq{jstar}.

\subsection{Convergence of the SCGF Estimator}\label{convergence}

When used with only a finite number $M$ of blocks of finite length $B$, the estimator $\hat{\tilde{\theta}}(k_{\delta q}, k_{a})$, defined in \eq{estimator}, can exhibit statistical and systematic errors. In this section, we analyze these errors to understand when the estimate of $J_1(\tilde{a}_0)$ calculated through the SCGF \eq{2D_scgf} will be accurate. As in the previous subsection, this analysis closely follows \cc{rohwer2015convergence}.
 
To quantify the statistical error associated with our calculated value of $J_{1}(\tilde{a}_0)$ we repeat the calculation procedure using $R$ independent trajectories. Each of these trajectories is split up into $M$ blocks of length $B$, and used to calculate $J_{1}(\tilde{a}_0)$. Our final estimate for the correction is then
\beq
\label{mean_estimator}
\hat{J}_1(\tilde{a}_0) = \frac{1}{R} \sum_{j=1}^R J_1^{(j)}(\tilde{a}_0),
\eeq
where $J_1^{(j)}(\tilde{a}_0)$ is the value of the correction calculated from the $j^{\rm th}$ trajectory. The statistical error of \eq{mean_estimator} can be estimated using
\beq
\label{error}
\mathrm{Err}[\hat{J}_1(\tilde{a}_0)] = \sqrt{\frac{\mathrm{Var}[\hat{J}_{1}(\tilde{a}_0)]}{R}}.
\eeq

This statistical error is only meaningful if we know that the systematic error in the calculation is comparatively small. There are two sources of systematic error that arise when using \eq{estimator}: correlation error and linearization error. Correlation error results from the fact that the derivation of the estimator assumes that the trajectory blocks are long enough to be approximately independent. This will be true if $B > T_{\rm corr}$ where $T_{\rm corr}$ is the correlation time of the reference model. If, however, the sub-sampled blocks of a trajectory are correlated, meaning that $B < T_{{\rm corr}}$, then we will not obtain an accurate estimate of $\tilde{\theta}(k_{\delta q}, k_{a})$ even as $M,R \to \infty$.

One way to resolve this correlation issue is to increase the block time $B$, but this also increases the magnitude of the other systematic source of error, linearization error. Linearization error is a manifestation of the fact that trajectories that contribute most to the average in the SCGF \eq{2D_scgf} for $k_a, k_{\delta q} \neq 0$ have atypical values of $\delta q$ and $a$. Using larger $B$ creates more self-averaging within a single trajectory and, as a result, makes sampling these atypical values more difficult. Linearization error also increases for fixed $B$ with increasing $|k_a|$ and $|k_{\delta q}|$, because larger values of these parameters weight rare trajectories' contributions to the SCGF more heavily. \cc{rohwer2015convergence} contains a detailed discussion of these problems. The authors of that work describe a method for checking to see if linearization error will substantially influence the estimate of the SCGF at a point $(k_{\delta q}, k_a)$ for some fixed $B$. We use this check, modified to account for the fact that our SCGF is two dimensional, as follows.

First, we calculate the SCGF \eq{estimator} along another 1D slice through its 2D domain. This slice is defined by fixing $k_{\delta q} = 0$ and increasing $k_{a}$, starting from $k_{a} = 0$~\footnote{The choice of moving in the positive $k_a$ direction is arbitrary. Moving in the negative $k_{a}$ direction will yield an equivalent convergence criterion (similarly we can move in either the positive or negative $k_{\delta q}$ directions while holding $k_a = 0$).}. By using finite difference along this slice we can calculate how $a$ varies with $k_a$. To compute the statistical error, we generate $R$ independent trajectories (usually the same $R$ trajectories we used to get \eq{error}), split each one into $M$ blocks of length $B$, and use each data set to calculate $a(k_a)$ along the slice. Our final estimate for the value of $a$ at each $k_{a}$ is
\beq
\hat{a}(k_{a}) = \frac{1}{R} \sum_{j=1}^R a^{(j)}(k_{a}),
\eeq
where $a^{(j)}(k_{a})$ is calculated from the $j^{\rm th}$ trajectory.

We will not end up using the $\hat{a}(k_a)$ values themselves. Instead, we focus on the associated values of the statistical error, calculated in the same way as the error of \eq{estimator},
\beq
\label{error_a}
\mathrm{Err}[\hat{a}(k_a)] = \sqrt{\frac{\mathrm{Var}[\hat{a}(k_a)]}{R}}.
\eeq
A plot of $\mathrm{Err}[\hat{a}(k_a)]$ as a function of $k_{a}$ will peak at some point $\hat{k}_{a}'(B)$ and then decline. Again $k_{\delta q} = 0$ is fixed during this entire calculation. As discussed in \cc{rohwer2015convergence}, $\hat{k}_{a}'(B)$ is an estimate for the maximum value $k_a'(B)$ at which the calculation of the SCGF will converge without being overwhelmed by linearization error. $k'_a(B)$ is a decreasing function of $B$, because linearization error grows as $B$ is increased.

Next we note that $\tilde{\theta}(k_{\delta q} = 0, k_{a}')$ corresponds to a point on the rate function $\tilde{J}(a')$, via the Legendre transform \eq{lt}. If we can converge the value of $\tilde{J}(a')$ then we can, with the same data set, also converge the value of $\tilde{J}(\delta q, a)$ at any point for which
\beq
\tilde{J}(\delta q, a) < \tilde{J}(a').
\eeq
This statement is intuitive in the context of probabilities and rate functions. However, it also applies when working with the SCGF, provided that $\tilde{J}(\delta q, a)$ is convex\c{chetrite2013nonequilibrium}. Thus the value of a point $\hat{\tilde{\theta}}(k_{\delta q}, k_{a})$ estimated using \eq{estimator} will be unaffected by linearization error if the associated point on the rate function, $\hat{\tilde{J}}(\delta q, a)$, satisfies
\beq
\label{convergence_check}
\hat{\tilde{J}}(\delta q, a) < \xi \hat{\tilde{J}}(a'),
\eeq
where $\xi < 1$ is an empirical constant (we set $\xi = 0.8$). The terms $\hat{\tilde{J}}$ in \eq{convergence_check} are averages over $R$ independent data sets of the corresponding Legendre transform \eq{lt}. This formula allows the convergence criteria derived in\c{rohwer2015convergence} for estimating one-dimensional SCGFs and rate functions to be applied in an arbitrary number of dimensions. 

We now discuss the procedure we use to converge the correction \eq{jstar} while accounting for correlation and linearization errors. For fixed block time $B$ we first increase $M$ and $R$ until the error bars for $\hat{J}_1(\tilde{a}_0)$, \eq{error_a}, are smaller than a desired value. We repeat this process for larger and larger $B$ until $\hat{J}_1(\tilde{a}_0)$ becomes independent of $B$. This is equivalent to increasing $B$ until it becomes larger than the reference-model correlation time $T_{\rm corr}$. If this happens while the convergence criterion \eq{convergence_check} holds at $\tilde{J}(\delta q^\star, \tilde{a}_0)$ then the calculation has succeeded, and we have computed (to within statistical error) the exact value of the correction $J_1(\tilde{a}_0)$.

If, however, the convergence criterion \eq{convergence_check} fails to hold in the regime in which \eq{estimator} still changes rapidly with $B$, then the bound $J_0(\tilde{a}_0)$ is too far from the exact answer for us to effectively sample $J_1(\tilde{a}_0)$ using direct simulation of the reference model. In this case the chosen ansatz is probably missing a crucial piece of the physics of the rare trajectories of the original model. On several occasions our failure to converge $J_1(\tilde{a}_0)$ based on an initial guess led us to construct a modified ansatz from which we could converge the exact correction. In the cases described in this paper we were able to reconstruct $J(a)$ using physically transparent ans\"atze containing only a modest number of parameters.

A quick way to estimate the scale of the correction is to compute the first term in \eq{cumu}, which requires computing only the variance of $\delta q$ within the reference model. By the central limit theorem we will have $\tilde{J}(\delta q,\tilde{a}_0) \approx (\delta q)^2/(2 T \sigma_{\rm ref}^2)$ close enough to the origin, where $\sigma_{\rm ref}^2 \propto 1/T$ is the variance of $q$ within the reference model. If $\sigma_{\rm ref}^2$ is small (which is the case when the ansatz is very good) then the correction $J_1(\tilde{a}_0) \approx T \sigma_{\rm ref}^2/2$. Thus if $ T \sigma_{\rm ref}^2/2$ looks small when plotted in the form of \f{fig0} it might be worth attempting to compute the correction \eq{j1}. If not, a better reference model ansatz is probably required.

\subsection{Efficiency of the correction calculation}
\label{efficiency}

Standard arguments are often used to suggest that computing the exact value of $J(a)$ is not possible without knowledge of the exact rare dynamics, or the use of methods such as cloning or transition-path sampling. This claim is based on the fact that computing $J(a)$ requires computation of the integral in \eq{int1}, and assumes that because the integrand grows exponentially with $T$, the number of trajectories required to converge this expression as $T \to \infty$ is prohibitively large. While this latter statement is correct, it does not speak directly to the task at hand. To compute the integral we do not need to take $T \to \infty$. Instead, we need $T \gtrsim T'$ where $T'$ is smallest time at which the large deviation principle applies. Taking $T \gg T'$ makes sampling more difficult and is unnecessary. It is possible for $T'$ to be large, but if the reference-model dynamics are close enough to the rare dynamics of the original model then the variance of the $q$ will be small and it will be possible to converge this integral term without using an unreasonable number of trajectories.

Moreover, computing the correction is numerically cheaper than inspection of the integral alone might suggest. If we switch to the SCGF representation, \eq{2D_scgf}, we can instead work with trajectories of length $T = T_{\rm corr}$, where $T_{\rm corr}$ is the correlation time of the reference model (in practice these trajectories are constructed by sub-sampling a single longer trajectory). Generally, $T_{\rm corr}$ is much smaller than $ T'$, the time required for the large-deviation principle to apply, and so sampling the moderately rare events required to reconstruct the auxiliary rate function $\tilde{J}(\delta q,a)$ near its minimum is cheaper in the SCGF representation. This property is ideal for the present method because we have reduced the problem of sampling $J(a)$ arbitrarily far from its minimum to one of sampling $\tilde{J}(\delta q, \tilde{a}_0)$ (potentially) close to its minimum. Working with the SCGF also removes the constraint $a = \tilde{a}_0$ present in \eq{j1}. Finally, we note that we are calculating the SCGF that is Legendre dual to the correction term \eq{j1}, and not the SCGF that is Legendre dual to the original rate function $J(a)$. Thus our method can in principle reproduce rate functions $J(a)$ that are not convex ($J_0$ is not required to be convex).

\subsection{Summary -- A variational ansatz for rare dynamics (VARD)} 

\begin{enumerate}
\item Given a continuous-time dynamics with rates $W_{xy}$ and a path-extensive dynamical observable $a$, we wish to determine $J(a)$, the large-deviation rate function for $a$ for trajectories of the model $W_{xy}$ of fixed time (assuming that $J(a)$ exists). We use a reference dynamics to calculate $J(a)$ as the sum of an upper bound $J_0(a)$ and a correction $J_1(a)$. The bound can always be calculated, and the correction can be calculated if the criteria described in \s{convergence} hold. If so then we succeed in calculating $J(a)$; if not, then the method returns an upper bound $J_0(a) \geq J(a)$.
\item Determine a reference-model dynamics \eq{three} and \eq{four} able to produce more or less of $a$ than the original model. In this paper we set the arbitrary bias $\beta_{xy}$ using physical intuition. 
\item Run a set of reference-model trajectories for different values of the reference-model parameters ($s$, $\lambda$, $\beta_{xy}$). For each trajectory, evaluate $\tilde{a}_0$ and $\tilde{q}_0$, using Eqs. \eq{ay} and \eq{kew}, and then use \eqq{j0} to plot the point $(\tilde{a}_0,J(\tilde{a}_0))$; see \f{fig0}(a). The lower envelope of these points over values of the reference-model parameter set is the tightest upper bound on $J(a)$ associated with the ansatz chosen in Step 1. 
\item Attempt to calculate the correction $J_1(a)$ at points on the bound by running a few ($\sim 5$) trajectories for each reference model. With the data from each trajectory, calculate $\tilde{J}(\delta q^\star, \tilde{a}_0)$, \eqq{eval_lt}, using Eqs. \eq{estimator}, \eq{calc_kta0} and \eq{calc_dq_star}. Insert the resulting values into \eqq{jstar}. Calculate the averaged correction $\hat{J}_1(a)$ and associated statistical error using Eqs. \eq{mean_estimator} and \eq{error}.
\item To verify convergence, repeat the calculation of Step 3 for increasing values of the block time $B$, until the averaged estimate for the correction \eqq{mean_estimator} no longer changes with $B$, {\em and} the convergence criterion \eq{convergence_check}, with $\delta q=\delta q^\star$ and $a=\tilde{a}_0$, holds~\footnote{The quantity $\hat{\tilde{J}}(a')$ on the right-side of \eq{convergence_check} is determined from the Legendre transform \eq{lt} of $\hat{\tilde{\theta}}(k_{\delta q} = 0, k_a = \hat{k}_a')$. The quantity $\hat{k}_a'$ is calculated by finding the maximum of $\mathrm{Err}[\hat{a}(k_a)]$, \eqq{error_a}, while starting at the point $(k_{\delta q} = 0, k_a = 0)$ on the SCGF and scanning outwards by increasing $k_a$.}. The accompanying GitHub script\c{jacobson2019vard} performs steps 4 and 5 automatically.
\end{enumerate}

\begin{figure*}[] 
   \centering
  \includegraphics[width=\linewidth]{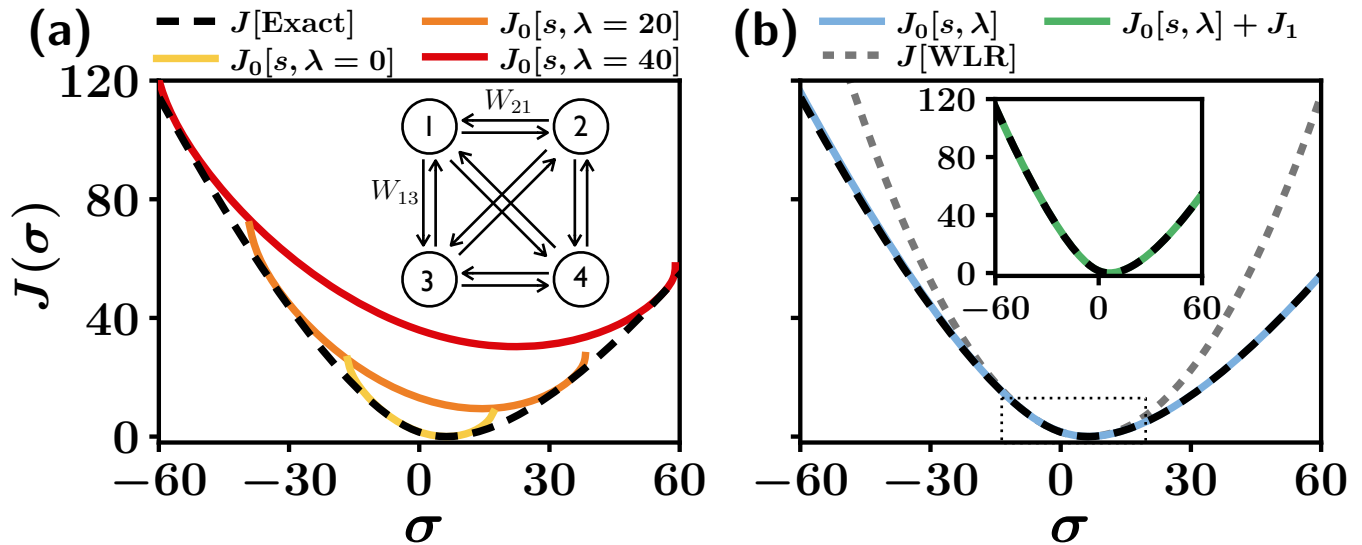} 
   \caption{Entropy production rate $\sigma$ for the 4-state model of \cc{gingrich2016dissipation}, the rates $W_{xy}$ of which are given by \eq{rates}. (a) Bounds \eq{j0} for the large-deviation rate function $J(\sigma)$. Each point on the colored lines results from a reference-model dynamics [Eqns.~\eq{three} and~\eq{four}] obtained by scanning the indicated parameters. The black dashed line is the exact answer. (b) Bound (blue line) resulting from a scan of the parameters $s$ and $\lambda$ compared with the WLR bound of~\cc{gingrich2016dissipation}, \eqq{wlr}. Inset: The sum (green line) of the bound and the correction \eq{j1} equals the exact rate function, arbitrarily far into its tails. The boxed region in panel (b) indicates the scale of Fig. 2 of \cc{gingrich2016dissipation}. Error bars for both axes are smaller than the thickness of the lines.}
   \label{fig1}
\end{figure*}

\section{VARD applied to four examples}
\label{app}

\subsection{Summary of the section} 

We now apply the method to four models taken from the literature. In each case, a simple and physically-motivated ansatz for the modified dynamics allows computation of the exact rate function $J(a)$. We focus on models whose state space is small enough that the exact rate function can be computed by standard methods -- Legendre transform of the SCGF calculated using the largest eigenvalue of the tilted rate matrix\c{touchette2009large} -- in order to validate our method (at the end of the section we also use VARD to compute descriptive rate-function bounds for two systems too large to solve by matrix diagonalization). In figures, the exact rate function is shown as a black dashed line. Absent the exact answer we would apply the method in exactly the same way. For a given reference model the fluctuations of the quantity $q$ reveal whether the bound $J_0(a)$ is tight, and whether we can compute $J(a)$ exactly. 

\subsection{Entropy production in a 4-state model}
\label{four_state}
We start by sampling entropy production in the 4-state model of \cc{gingrich2016dissipation}, shown in the inset of \f{fig1}(a). This is a fully-connected network model with transition rates 
\bea
\label{rates}
\begin{array}{lll} 
    W_{12} = 3,& W_{13} = 10,& W_{14} = 9,\\
    W_{21} =10,& W_{23} = 1,& W_{24}  = 2,\\ 
    W_{31} = 6,& W_{32} = 4,& W_{34} =1,\\
W_{41} = 7,& W_{42} = 9,& W_{43} = 5.
   \end{array}
   \eea
These rates do not satisfy detailed balance, and so the model produces nonzero entropy on average. To quantify the fluctuations of entropy production for trajectories of fixed time we construct a reference model as follows. The dynamical observable is $a=\sigma =T^{-1} \sum \alpha_{xy}$, where the sum is taken over all jumps $x \to y$ of a trajectory, and $\alpha_{xy} = \ln (p_{xy}/p_{yx})$, where $p_{xy} = W_{xy}/\sum_y W_{xy}$. Our basic reference-model parameterization is defined by the parameters $s$ and $\lambda$ appearing in \eq{three} and \eq{four}, together with any additional set of biases $\beta_{xy}$ suggested by the physics of the problem under study. Here we reason that none is necessary: the bias $\lambda$ is always required, in order to sample jump times, and the bias $s$ is sufficient to influence the entropy produced per jump, $A/K$ (a fact that is easy to guess, and to confirm with some short simulations). We therefore impose no additional bias, and set $\beta_{xy}=0$. 

We ran trajectories for a fixed number $K = 1.5 \times 10^8$ of events, roughly equivalent to a time of $T=10^7$ in the unbiased model. We simulate in the constant-event ensemble for convenience, because there all simulations, regardless of the value of $\lambda$, take approximately the same amount of processor time. The equations of \s{desc} then allow us to compute the rate function for the original model in the constant-{\em time} ensemble (see \cc{budini2014fluctuating} for more on the relationship between the constant-event and constant-time ensembles).

The bounds $J_0(a)$ resulting from a scan of $s$, for three values of $\lambda$, are shown as colored lines in \f{fig1}(a). In figures we use the compact notation $J_0[x]$ to indicate the bound $J_0(a)$ swept out by scanning the set of parameters $\{x\}$. We also show the exact rate function (black dashed line), obtained by matrix diagonalization. Different combinations of $s$ and $\lambda$ produce the best (lowest) bound at different values of $a$, so indicating the ``least unlikely way'' of realizing each value of $a$ within the manifold of dynamics accessible to the reference model. The bound produced by the scan $\lambda=0$ provides the best bound close to the typical value $a_0$, but not far from it, indicating that very rare values of $a$ are produced by the original model using a combination of rare jump types ($s\neq0$) {\em and} rare jump times ($\lambda \neq 0$).

The bound swept out by scanning both $s$ and $\lambda$ is shown in blue in \f{fig1}(b). We used $201$ equally spaced $s$ values on the interval $[-5, 5]$, and $51$ equally spaced $\lambda$ values on $[0, 50]$. This bound lies close to the exact answer, even far into the tails of the rate function. For comparison we show the weakened linear response (WLR) universal current bound of \cc{gingrich2016dissipation} (gray dashed line); the dotted box in the center of the figure indicates the scale of Fig. 2 of that paper. The WLR bound is
\beq
\label{wlr}
J[{\rm WLR}]= \frac{\sigma_0}{4 c_0^2} (c-c_0)^2,
\eeq
where $c$ is a current, $c_0$ is its typical value (in the original model), and $\sigma_0$ is the typical value of the rate of entropy production. $c_0$ and $\sigma_0$ must be determined by running a single trajectory of the original model, and \eq{wlr} then provides a bound on the probability of observing an atypical value of $c$. The bound is tightest in the case $c \propto\sigma$. 

The WLR bound can be derived from Level 2.5 of large deviations\c{maes2008canonical,bertini2015large} using a mean-field ansatz that assumes all currents scale with time in the same way (for both forward and time-reversed versions of the model). By contrast, the $(s,\lambda)$-bound results from a microscopic ansatz, \eq{three} and \eq{four}, inserted into the exact result \eq{av} for the dynamical partition sum, and does not assume that all currents scale in the same way. Inspection of the tails of the bounds reveals that the microscopic ansatz captures the rare behavior of the model more accurately than does the homogeneous ansatz. Thus we learn that the rare behavior of even this very simple model does not simply resemble a speeded-up or slowed-down version of its typical behavior. The bound \eq{wlr} is a universal statement about the physics that constrain fluctuations, and is not designed to be a means of accurate numerical sampling. Nonetheless, it is meaningful and instructive to compare the bounds produced by different types of ans\"atze. 

For each of the reference models that lie on the bound $J_0[s,\lambda]$ we calculate the correction \eq{j1} using the procedure described in \s{corr}. For all points we obtain convergence of the correction. The result, $J_0(a)+J_1(a)$, is shown as a green line in the inset of \f{fig1}(b), and matches the exact answer (black), obtained by matrix diagonalization, as it should.  We used $10^4$ blocks, each of length of $50T_{\rm event}$, where $T_{\rm event}$ is the time per event for each reference model. The average correction \eq{mean_estimator} and statistical error are obtained from $10$ independent data sets. Error bars on the rate function are calculated by combining the error from the correction and the error from the bound according to
\beq
\mathrm{Err}[\hat{J}(\tilde{a}_0)] = \sqrt{(\mathrm{Err}[\hat{J}_0(\tilde{a}_0)])^2 + (\mathrm{Err}[\hat{J}_1(\tilde{a}_0)])^2}.
\eeq
The error in the bound is estimated by running $10$ additional trajectories at each point and calculating the standard deviation of $J_0(\tilde{a}_0)$. The standard deviation of $\tilde{a}_0$, calculated in the same way, yields error bars for the horizontal axis.

\begin{figure*}[] 
   \centering
   \includegraphics[width=\linewidth]{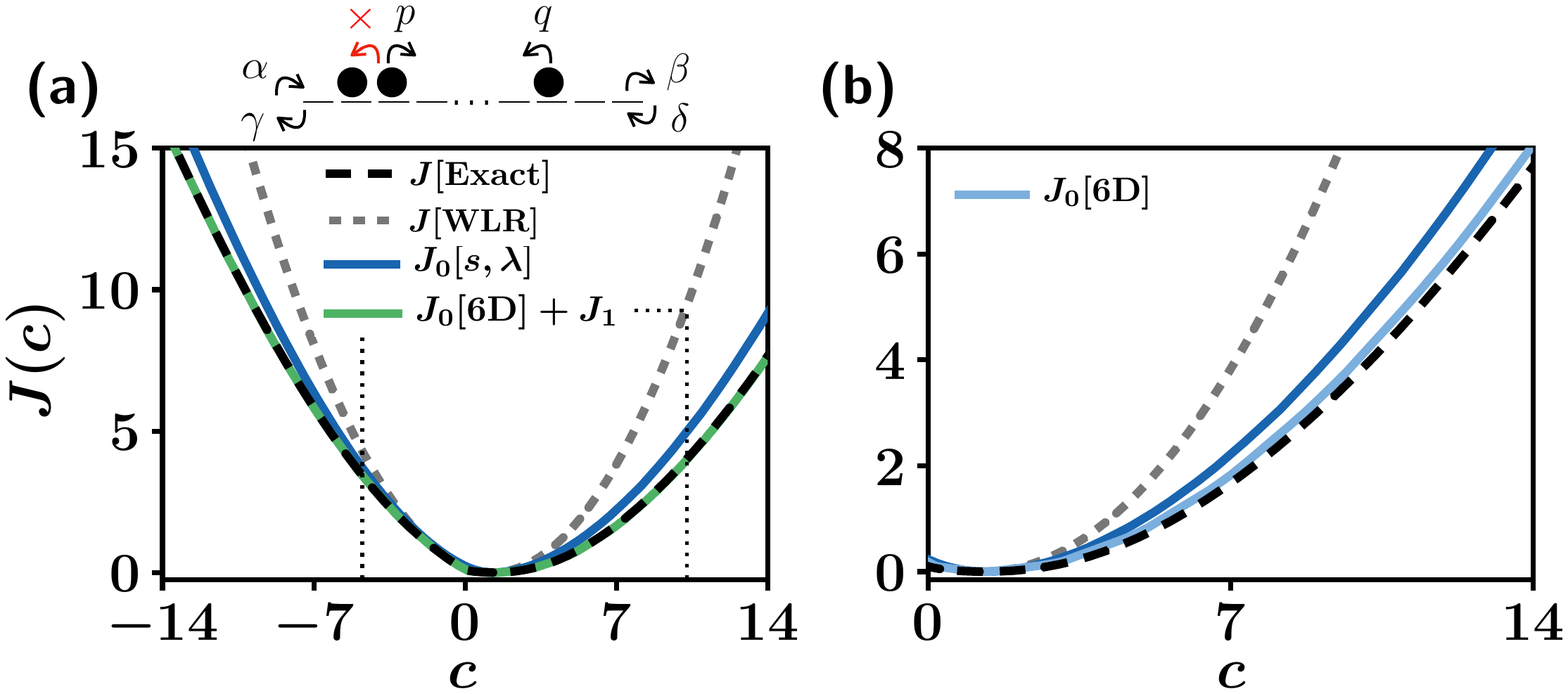} 
   \caption{Large-deviation rate function $J(c)$ of particle current, $c$, for  the version of the ASEP studied in~\cc{gingrich2016dissipation}; rate constants are given in the text. (a) We show the exact answer (black), the WLR universal current bound\c{pietzonka2016universal,gingrich2016dissipation} (gray), and the bound produced using our default $(s,\lambda)$ reference-model parameterization (dark blue). (b) With additional physical insight used to refine the reference model it is possible to tighten the bound (compare dark blue and light blue lines). The green line in (a) is the corrected version of the 6D bound. The boxed region in panel (a) indicates the scale of Fig. 3 of \cc{gingrich2016dissipation} (not the scale of \f{fig2}(b)). Error bars for both axes are smaller than the thickness of the lines.}
   \label{fig2}
\end{figure*}

From this example it is clear that VARD is numerically much more efficient than direct simulation (of the original model): accurate calculation of the rate function at values of order 100 indicates accurate calculation of probabilities of order $\e^{-100 T}$, where $T$ is the elapsed time of the trajectory. We do not know in advance which values of reference-model parameters constitute good choices, for particular values of $a$, but it is a simple matter to scan parameters and pick the smallest value of $J_0$ given $a$. Additional sampling then allows us to determine if we can calculate the correction \eq{j1}, and therefore the exact rate function. We were able to do this here with little additional numerical effort. In this example, the state space of the model is small enough that its dynamics can be solved by matrix diagonalization, and so we possess the exact answer in advance. We made this choice because we wish to benchmark the method. However, our procedure would be identical if we did not know the exact answer ahead of time: define the reference model, pick the best bound, and attempt to calculate the correction term. The results of the latter calculation tell us if we have the exact answer, or, if not, roughly how close we are to obtaining it. If we are not close at all then we need a better reference-model guess. Inspection of the bounds produced by different reference models is also physically instructive, indicating the extent to which certain dynamical mechanisms contribute to the rate function at particular values of the order parameter.

\subsection{Current in the ASEP}\label{asep}
We next sample current in the asymmetric simple exclusion process (ASEP), a model of hard particles that hop between lattice sites\c{derrida1998exactly,chou2011non} (an interesting alternative would be to consider the symmetric simple exclusion process, which has fewer parameters but also shows complex scaling behavior\c{derrida2002large}). We consider the version of the model studied in Fig.~3 of\c{gingrich2016dissipation}, shown in \f{fig2}, with open boundaries and a lattice of $L=15$ sites. The rate constants are $\alpha = 1.25, \beta = 0.5, \gamma = 0.5, \delta = 1.5, p = 1, q = 0.5$, placing the model in the high-density region of the ASEP phase diagram\c{kolomeisky1998phase,blythe2000exact}. The dynamical observable is $a=c =T^{-1} \sum \alpha_{xy}$, where the sum is taken over all jumps $x \to y$ of a trajectory, and $\alpha_{xy} = \pm 1$ if the jump $x \to y$ sees a particle move to the right (upper sign) or left (lower sign).

In \f{fig2}(a) we show the bound swept out by our default $(s,\lambda)$ reference-model parameterization (dark blue), which provides a meaningful numerical bound on the exact rate function (black) even far into the tails. We ran trajectories for $K=5 \times 10^5$ events, roughly equivalent to a time $T=10^5$ in the unbiased model. We scanned $81$ equally spaced $s$ values on the interval $[-2, 2]$, and $41$ equally spaced $\lambda$ values on $[0, 10]$. Also shown is the WLR universal current bound\c{pietzonka2016universal,gingrich2016dissipation} (gray). The WLR bound describes accurately the moderately rare behavior of the model, but not the very rare behavior, which is quantified by the tails of the rate function. Comparison of bounds indicates, as in the previous subsection, that very rare currents are generated by trajectories that do not resemble speeded-up or slowed-down versions of the forward- or backward-running typical dynamics: the configurations visited in the tails of the rate function are different to those visited near the center. 

While the ($s, \lambda$)-bound is already meaningful, it is possible to produce tighter numerical bounds by guessing additional ways in which the very rare high- or low-current behavior might be achieved. Inspection of the way in which $s$ couples to the rate constants (here any rate involving a hop to the right is multiplied by $\e^{-s}$ and any rate involving a hop to the left is multiplied by $\e^{s}$) reveals that varying $s$ moves the reference model around the ASEP phase diagram\c{kolomeisky1998phase,blythe2000exact}. The original model sits in the high-density region of phase space, but the reference model need not. Inspection of the phase diagram indicates that the end rates $\alpha, \beta,\gamma,\delta$, separate from the bulk rates $p$ and $q$, play a key role in determining the ASEP's typical behavior: if particles are fed in relatively quickly or slowly then we reside in the high- or low-density region of phase space, respectively, and if input- and output rates are balanced then we can access the maximum-current region. Returning to \eq{rt} we introduce a set of parameters $\beta_{xy}$ that couple to the end rates, such that the rate $\alpha$ in the original model becomes $\e^{-u_\alpha} \alpha $ in the reference model ($u_\alpha$ being a parameter), and similarly for the three other end rates. We also include a contribution to $\beta_{xy}$ that biases trajectories toward or away from creating particle-particle contacts (i.e. particles on adjacent sites), reasoning that controlling such contacts can help control the escape rate of visited configurations, so helping cause or prevent traffic jams. A simple way to do this is to add to $\beta_{xy}$ a bias $-\mu \Delta_{xy}$, where $\mu$ is a parameter and $\Delta_{xy}$ is the change in the number of particle-particle contacts when moving from $x$ to $y$. 

With the new bias determined we can generate an improved bound for the ASEP.  We split the calculation into two parts and focus separately on the piece of the rate function for values of the observable greater than the mean, $a > a_{0}$, and less than the mean, $a < a_{0}$. Since the dynamics in these two different regimes are qualitatively different, generating an effective set of reference models for each requires scanning over different regions of the ansatz parameter space. For $a < a_0$ we scanned $s, \lambda, u_{\gamma}$ and $\mu$. For $a > a_{0}$ we scanned $s, \lambda, u_{\alpha}, u_{\beta}$ and $\mu$,  making $6$ parameters in total. Combining these calculations produces the 6D bound shown in light blue in \f{fig2}(b). This bound is tighter than the default $(s,\lambda)$-bound. The 6-parameter scan can be carried out with reasonable numerical effort: for a given set of parameters we need only a short single trajectory to compute the averages required for the bound, and there is no requirement for communication between the calculations. On the left side we scanned $11$ equally spaced $s$ values on the interval $[0, 2]$; $21$ equally spaced $\lambda$ values on the interval $[-0.6, 0]$, and $17$ more on the interval $[0, 8]$; $29$ $u_{\gamma}$ values so that $\gamma$ takes on equally spaced values on the interval $[0.1, 1.5]$; and $31$ equally spaced $\mu$ values on the interval $[-1, 0.5]$. On the right side we scanned $11$ equally spaced $s$ values on the interval $[-2, 0]$; $17$ equally spaced $\lambda$ values on the interval $[0, 8]$; $21$ $u_{\alpha}$ values and 21 $u_\beta$ values so that $\alpha$ and $\beta$ each take on values equally spaced on the interval $[0.5, 1.25]$; and $11$ equally spaced $\mu$ values on the interval $[0, 0.5]$. The lower envelope of the values of \eq{j0} constitutes the improved bound.

Correcting the 6D bound by calculating the correction $J_1$ at points along the bound gives the green line shown in \f{fig2}(a), which agrees with the exact rate function even far into the tails. For this calculation we used $10^4$ blocks of length $100T_{\rm event}$, where $T_{\rm event}$ is the time per event in each reference model. Errors are computed as in Section \ref{four_state}.

\begin{figure}[] 
   \centering
  \includegraphics[width=\linewidth]{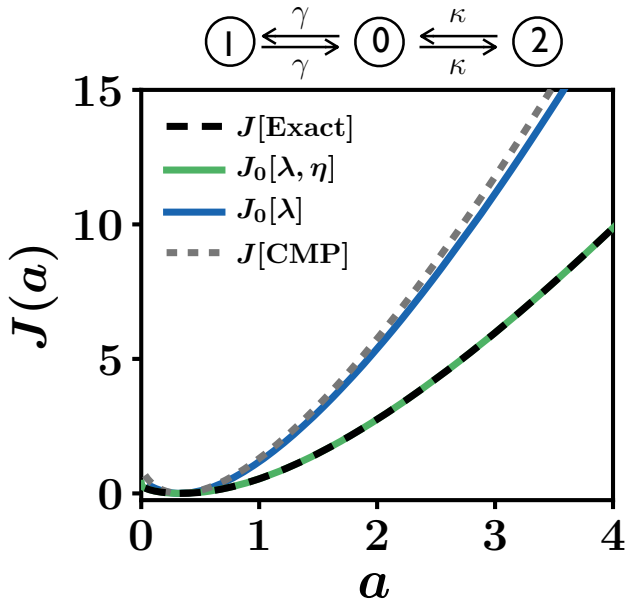} 
   \caption{Large-deviation rate function $J(a)$ for the number of jumps $1 \to 0$ per unit time, $a$, for the 3-state model of \cc{garrahan2017simple}. We compare with the exact answer (black) the bounds produced by reference models in which we control jump times (blue), or jump times and jumps from $0 \to 1$ (green). Error bars for both axes are smaller than the thickness of the lines. Also shown is the CMP universal activity bound\c{garrahan2017simple}, \eqq{cmp} (gray).}
   \label{fig3}
\end{figure}

\subsection{Activity in a 3-state model} We consider the three-state model of Fig. 3 of \cc{garrahan2017simple}, shown in \f{fig3}. The rate constants are $\gamma=1$ and $\kappa=1/2$. Our chosen dynamical observable, $a$, is the number of jumps from states $1 \to 0$ per unit time. The parameter $s$ in our default $(s,\lambda)$ reference-model parameterization \eq{rt} has no role to play here: $s$ controls the probability of the $1 \to 0$ process, but once in state 1 there is nowhere to go but state 0. Thus $s$ cannot influence the number of counted events per jump, $A/K$, and so we set $s=0$. 
\begin{figure*}[] 
   \centering
  \includegraphics[width=\linewidth]{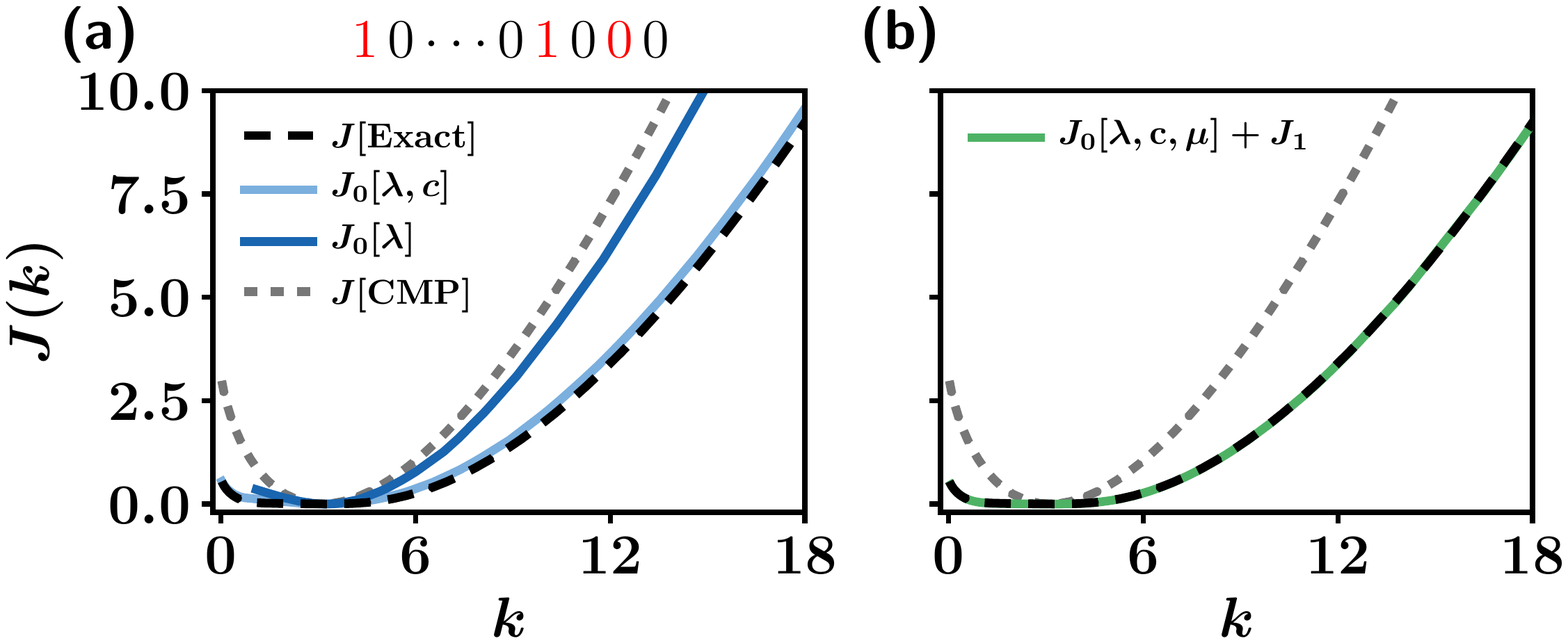} 
   \caption{Large-deviation rate function $J(k)$ for the number of jumps per unit time, $k$, for the one-dimensional Fredrickson-Andersen model. (a) We compare with the exact answer (black) the bounds produced by reference models in which we control jump times (dark blue) or jump times and the mean up-spin fraction (light blue). Also shown is the CMP universal activity bound\c{garrahan2017simple}, \eqq{cmp} (gray). (b) The sum (green) of the bound (produced using a reference model in which we control jump times, mean up-spin fraction, and pair correlations) and the correction matches the exact answer.  Error bars for both axes are smaller than the thickness of the lines.}
   \label{fig4}
\end{figure*}

At this point we need to apply our physical intuition in order to create a reference-model ansatz able to control $A/K$. Inspection of the network reveals that controlling the jump destination from state 0 is sufficient for this purpose: if we jump $0 \to 1$ then we must subsequently jump $1 \to 0$, whereas if we jump $0 \to 2$ we will return to 0 without making the counted jump. In \eqq{rt} we therefore set $\beta_{01}=\eta$ (a parameter) such that the reference-model rate for the process $0 \to 1$ is $\tilde{W}_{01} = \e^{-\eta} \gamma$. We set all other $\beta_{xy}=0$. Scanning $\eta$ and $\lambda$ (our usual jump-time bias) produces the bound shown in green in \f{fig3}. Bounds were calculated using $11$ equally spaced $\lambda$ values on the interval $[-0.5, 0]$ and $101$ values on the interval $[0, 25]$, and $51$ equally spaced $\eta$ values on the interval $[-5 , 5]$. All trajectories were run for $K=10^7$ events, roughly equivalent to a time of $T=10^7$ in the unbiased model. Error bars are computed as in Section \ref{four_state}.

The $(\eta,\lambda)$-bound is effectively exact, as we can deduce by measuring the fluctuations of $q$ (which here are nonexistent). In this case the model is simple enough that each reference model used to compute the bound enacts the exact rare dynamics of the original model, conditioned on a particular value of $a$. As a result, the correction term $J_1$ vanishes, and the bound $J_0$ is exact. (This exactness is reasonable on account of the fact that the system has relatively few ways of realizing values of $A/K$ and $K/T$, but it is not obvious in advance that the chosen parameterization would require no additional correction.) Recall that the correction term can be interpreted as a measure of how close the typical dynamics of the reference model is to the desired rare dynamics of the original model; here, typical trajectories of the reference model correspond exactly to trajectories of the original model conditioned on the relevant value of the order parameter. 

The chosen observable is a non-decreasing counting variable, not a current, and so the universal bound of Refs.\c{pietzonka2016universal,gingrich2016dissipation} does not apply. One that does apply is the Conway-Maxwell-Poisson (CMP) bound of \cc{garrahan2017simple},
\beq
\label{cmp}
J[{\rm CMP}] = \frac{k_0}{a_0} \left( a \ln \frac{a}{a_0}+ a_0 -a\right),
\eeq
where $a$ is the dynamical observable, $a_0$ is its typical value (in the original model), and $k_0$ is the typical dynamical activity (the total number of events per unit time) of the original model (note that there is an $a$ missing in front of the logarithm in Eq.~(17) of\c{garrahan2017simple}). 

The CMP bound is shown in gray in \f{fig3}. Similar to the universal current bound, the CMP bound is derived from Level 2.5 of large deviations using an ansatz that assumes the rare behavior of the system to be a speeded-up or slowed-down version of its typical behavior. It therefore has similar properties to our $\lambda$-bound, shown in blue in \f{fig3} (the $\lambda$ bound is constructed from the pieces of the $(\eta,\lambda)$-bound with $\eta=0$). Comparison of this bound and the $(\lambda,\eta)$-bound shows the extent to which the very rare behavior of this model is dominated by trajectories comprising rare jump times {\em and} an atypical propensity to jump left from state 0. Analogous to its current counterpart, the CMP bound is a general statement about the physics controlling the fluctuations of counting variables, and is not intended to be a means of numerical sampling. Nonetheless, comparison of its properties with bounds obtained by the microscopic ansatz used here is instructive, and addresses the point raised in \cc{garrahan2017simple}: ``It would be interesting to find alternative yet simple variational ansatzes that can capture [the] strong fluctuation behavior [of the 3-state model]''.

\subsection{Activity in the FA model}
\label{sec_fa}

We consider the one-dimensional Fredrickson-Andersen (FA) model with periodic boundary conditions\c{fredrickson1984kinetic}. This is a lattice model with simple thermodynamics and with dynamical rules that give rise to slow relaxation and complex spatiotemporal behavior\c{garrahan2002geometrical}. On each site of a lattice (here of length $L=15$) lives a spin, which can be up or down. Up-spins (resp. down-spins) can flip down (resp. up) with rate $1-c$ (resp. $c$) if at least one of their neighboring spins is up; if not, then they cannot flip (the rate $c$ here should not be confused with the symbol for current in previous sections). In \f{fig4}(a), top, we show an example FA model configuration, with periodic boundary conditions; the spins in red cannot flip. Our chosen dynamical observable, $a=k=T^{-1} \sum \alpha_{xy}$, is the number of jumps per unit time, where $\alpha_{xy} =1$ for all jumps $x \to y$. The large-deviation properties of $k$ have been studied in detail, and give rise, in certain limits, to singular behavior in the SCGF that is Legendre dual to $J(k)$\c{garrahan2007dynamical,garrahan2009first}.
 \begin{figure*}[] 
   \centering
  \includegraphics[width=\linewidth]{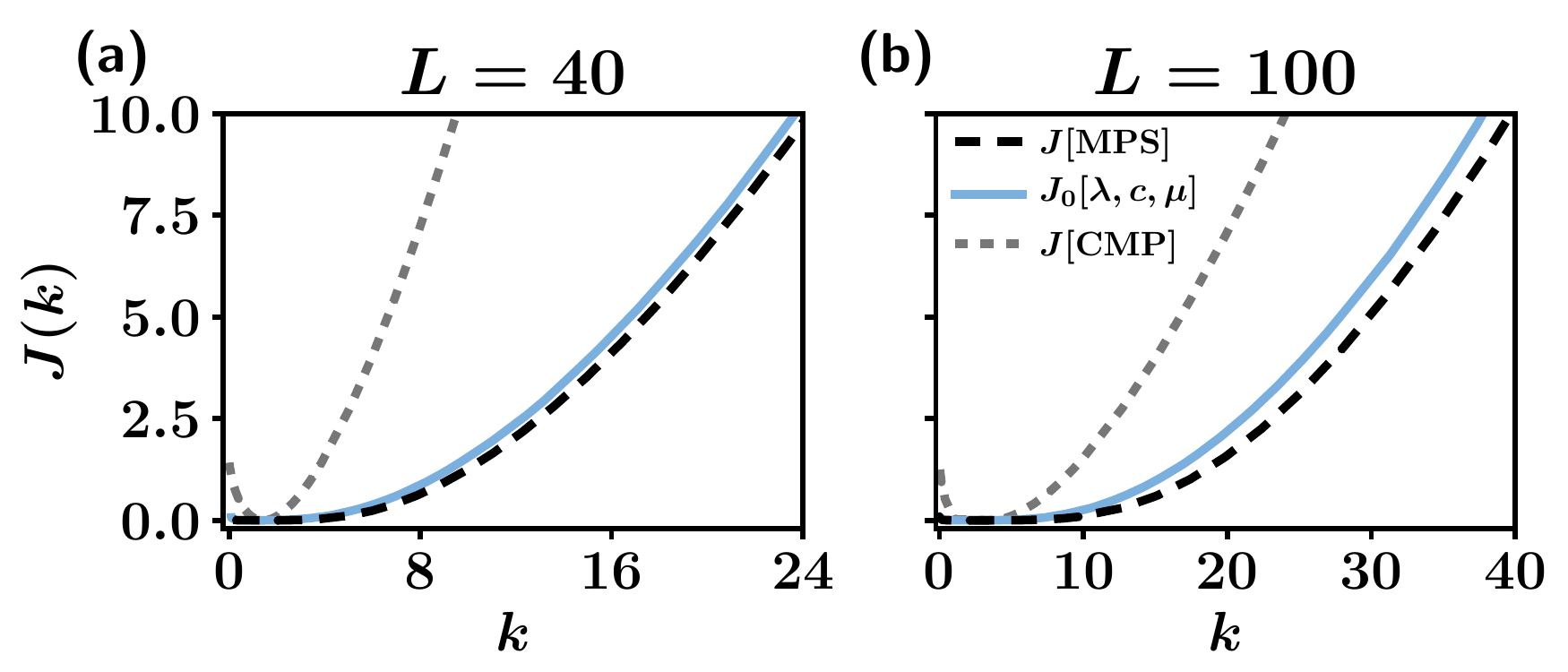} 
   \caption{Large-deviation rate function $J(k)$ for the number of jumps per unit time, $k$, for the Fredrickson-Andersen model of \cc{banuls2019using} with $c = 0.1$. Lattice sizes are (a) $L=40$ sites and (b) $L=100$ sites. We compare the matrix product state (MPS) calculation of \cc{banuls2019using} (black) with the three-parameter VARD bound of \f{fig4} (blue). Also shown is the CMP universal activity bound\c{garrahan2017simple}, \eqq{cmp} (gray). }
   \label{fig5}
\end{figure*}

In \f{fig4}(a) we show the CMP universal activity bound\c{garrahan2017simple} on $J(k)$ and the bound produced by our reference-model $\lambda$-scan. These are of similar character, because they assume that the rare behavior of the model is a speeded-up or slowed-down version of its typical behavior. All trajectories were run for $K=3 \times 10^5$ events, roughly equivalent to a time of $T=10^5$ in the unbiased model. We used $61$ equally spaced $\lambda$ values on the interval $[-0.6, 0]$, and $61$ more on the interval $[0, 12]$. To produce a tighter bound we need to assume that the configurations visited by rare trajectories are different to those visited by typical ones (the CMP bound assumes that they are the same). The parameter $s$ in our default ($s,\lambda$) reference-model parameterization \eq{rt} again has no role to play, because biasing all jumps equally is equivalent to biasing none. A simple alternative is to choose the bias $\beta_{xy}$ so that the reference model can generate a larger or smaller number of up-spins than is typical in the original model. We choose the parameters $\beta_{xy}$ so that the parameter $c$ in the original model becomes $\e^{-\eta} c$ in the reference model, with $\eta$ being a parameter. A $(\lambda,\eta)$-scan of the reference model produces the bound $J_0[\lambda,c]$ shown in light blue in \f{fig4}(a). This bound provides a reasonable approximation of the exact answer over the whole range of $k$, indicating that much of the physics of rare activity fluctuations of the FA model can be accounted for by considering the typical behavior of versions of the model with different values of the parameter $c$. Here we chose $\eta$ so that $c$ takes on $100$ equally spaced values on the interval $[0.02, 2]$. 

It is possible to tighten this bound by reasoning that there must exist spatial correlations between up-spins if we condition trajectories upon activity per unit time $k$. For instance, given an up-spin fraction of exactly $1/2$, the escape rate (the sum of rates leading out of a given state, a quantity relevant to the number of jumps per unit time) is maximized by having pairs of up-spins separated by pairs of down-spins, and minimized by having up-spins and down-spins alternate. We induce these types of spatial correlations using the same $\mu$-bias used for the ASEP in Section \ref{asep}, favoring more or fewer contacts between up spins. Scanning $c, \eta,$ and $\mu$  produces a bound slightly tighter than $J_0[\lambda,\eta]$ (not shown). We used $41$ equally spaced $\mu$ values on the interval $[-1, 1]$. From this bound we compute the correction $J_1$, and the sum of the bound and correction matches the exact answer: see \f{fig4}(b). We computed the correction by splitting the calculation into two pieces, one on either side of the mean value $a=a_0$. For $a < a_0$ we used $10^4$ blocks of length $50 T_{\rm event}$, where $T_{\rm event}$ is the average time per event in each reference model. For $a>a_0$ we used $10^5$ blocks of length $600T_{\rm event}$. The different block lengths generated by the convergence procedure (see \s{convergence}) for $a < a_0$ and $a > a_0$ signal that the correlations present in the dynamics are qualitatively different in each of these regimes. Understanding the nature of these correlations is of physical interest\c{garrahan2002geometrical,garrahan2009first}. Errors are calculated as in \s{four_state}.

\subsection{Toward large-scale calculations}
\label{large}

In this paper we have demonstrated proof-of-principle of VARD using network systems or lattice models whose state space is small enough that their rate functions can be obtained by matrix diagonalization, so providing a benchmark for the method. VARD can also be applied to systems too large for matrix diagonalization to be feasible, in order to produce bounds or (if the ansatz is good enough and convergence of the correction is obtained) exact rate functions. An active line of research is to study large versions of certain lattice models in order to determine how their large-deviation properties change with system size\c{garrahan2009first,nemoto2017finite,banuls2019using}. In these regimes, specialized techniques are necessary. For instance, in \cc{nemoto2017finite}, a cloning procedure combined with feedback control was used to calculate large-deviation functions for an FA model of $L=36$ sites. In \cc{banuls2019using}, a matrix product state (MPS) calculation was used to compute large-deviation functions for an FA model of sizes of order $L=100$ sites (these results show some differences with the results of \cc{nemoto2017finite}, indicating that this is a technically challenging regime). Note that the FA model of \cc{banuls2019using} has open boundaries and slightly different facilitation rules than used in the previous section: spins facilitated by two spins flip at twice the rate of spins facilitated by one spin.

In \f{fig5} we compare the MPS calculation of \cc{banuls2019using} with the three-parameter VARD bound used in \f{fig4} for two lattice sizes that are considered large by current standards (the parameter $c = 0.1$). In both cases the VARD bound is descriptive, capturing the main features and the trends with $k$ of the MPS result. The bound is less tight for the larger system size, suggesting that more terms in the ansatz are required as the system becomes larger. However, the bound quality, even using an ansatz containing only 3 parameters, remains reasonable. As for the ASEP, the natural next step is to include additional parameters in the ansatz in order to tighten the bound and calculate the correction (compare dark blue and light blue lines in \f{fig2}(b)). A natural way to develop improved bounds is to use Monte Carlo learning procedures in order to optimize reference models containing a potentially large number of parameters\c{whitelam2019evolutionary}.

\section{Conclusions}
\label{conc}

We have described how direct simulation of a variational ansatz for rare dynamics (VARD) can be used to compute bounds for large-deviation rate functions in continuous-time Markov chains~\footnote{The method works for Markov chains in discrete time, upon replacing the distributions \eq{two} and \eq{four} by $\delta(\Delta t-1)$; see earlier versions of the method\c{klymko2017rare,whitelam2018sampling,whitelam2018multi}.}. This approach requires only direct simulation of versions of the original model with modified rates, and so is technically simple and easy to implement. It is also physically instructive, in the sense that the quality of the bounds produced by different physical ans\"atze reveal the extent to which different types of dynamical processes contribute to the rare behavior of the model of interest.

If the ansatz is chosen well then bounds can be corrected to produce the exact rate function, arbitrarily far into its tails; in the literature it is often assumed or stated that such precision is not accessible via direct reweighting of trajectories, and requires the use of specialized numerical techniques such as cloning or path sampling. For the models studied in Figs.~\ref{fig1}--\ref{fig4}, two network models and two lattice models taken from the literature, it is possible to calculate the exact rate function using only simple and approximate guesses about the nature of the rare dynamics. Although this rare behavior can be complex, we are rarely working in the dark: the model itself can exhibit different behavior in different parameter regions, and often its rare behavior at one point in parameter space is similar to its typical behavior at another point in parameter space. For example, we have studied the ASEP in its high-density region, where (typically) the lattice is crowded and particles move slowly. The ansatz we used to calculate its current rate function is equivalent to guessing that the rare, high-current behavior in the high-density region is similar to the typical behavior in the maximum-current region, where particles move quickly and possess spatial anticorrelations. Similarly, the FA model is complex, but the likelihood of its rare behavior at one value of the parameter $c$ can be well approximated by looking at the typical behavior of models at different values of $c$. We have also shown that bounds that are descriptive in small systems remain so in systems too large to solve by matrix diagonalization; we will discuss this regime further in forthcoming work.

VARD is similar to classical umbrella sampling\c{torrie1977nonphysical,tenWolde_1998} in the sense that the rate function $\tilde{J}(a)$ of the reference model can be regarded as a nonequilibrium umbrella potential, concentrating sampling at a desired point: see \f{fig0}. It is different, however, in that VARD does not require overlapping sampling windows -- reference models are used independently --  and we compute an absolute rate-function value $J(a)$, as opposed to a free-energy difference. This latter distinction results from the fact that the path weight appearing in \eq{av} is known exactly, and at the sampling point $\tilde{a}_0$ we know that the rate function of the reference model vanishes; by contrast, in the equilibrium case we know the probability of visiting a certain state only up to a normalization constant, and we do not know the absolute free energy of the reference model (unless it is particularly simple\c{frenkel1984new}).

VARD provides insight into the approaches used to produce universal rate-function bounds from Level 2.5 of large deviations, via homogeneous ans\"atze\c{pietzonka2016universal,gingrich2016dissipation,garrahan2017simple}, by showing how relaxing such assumptions leads to the tightening of bounds in different sectors of parameter space. It is also complementary to numerical large-deviation methods that use path-sampling, cloning, or adaptive methods to calculate the SCGF~\footnote{One advantage of bypassing the SCGF and calculating the rate function directly, as we do here, is the ability to reconstruct rate functions that are not strictly convex\c{touchette2009large,klymko2017rare,nickelsen2018anomalous}.} that is Legendre dual to $J(a)$\c {giardina2006direct,touchette2009large,garrahan2009first,chetrite2015variational,jack2015effective,ray2018exact,nemoto2017finite,ferre2018adaptive}. Sometimes path sampling or cloning are used in isolation, and sometimes they are combined with a modified dynamics. VARD lies at the other extreme of the methods spectrum in the sense that it uses {\em only} a modified dynamics. The bounds that result provide a natural starting point for those specialized methods, because the set of reference models that live on the bounds already resemble the rare dynamics of interest. Indeed, direct sampling of those reference models is, in the cases described in Figs.~\ref{fig1}--\ref{fig4}, sufficient to recover the statistics (the constituent configurations and jump times) required to compute $J(a)$ exactly.

There are several possible variants of the present method. The $\lambda$-scan accesses roughly the same information as the universal bounds of Refs.\c{pietzonka2016universal,gingrich2016dissipation,garrahan2017simple}, and one possible numerical simplification would be to eliminate the $\lambda$-scan in favor of the universal bounds \eq{wlr} or \eq{cmp}. We have also simply scanned parameters in order to identify the best bounds associated with a given ansatz, but as the complexity of an ansatz grows it is be natural to replace the scan with an evolutionary Monte Carlo procedure\c{whitelam2019evolutionary}. For instance, consider a set of reference models having $N$ parameters, and construct an initial $(s,\lambda)$-bound. Take models at various points on this bound, and define a window $\Delta a$ for the observable. For each model, perturb the $N$ parameters, generate a short trajectory, and calculate \eqq{j0}. If this value is less than the current bound (and $a$ lies within the designated window) accept the new reference model; otherwise, retain the original.

\section{Acknowledgments} We thank Hugo Touchette, Tom Ouldridge, and Juan Garrahan for discussions and comments on the manuscript, and thank Juan Garrahan, Todd Gingrich, and Mari Carmen Ba{\~n}uls for providing data from Refs.\c{garrahan2017simple}, \c{gingrich2017fundamental}, and \c{banuls2019using}, respectively. This work was performed as part of a user project at the Molecular Foundry, Lawrence Berkeley National Laboratory, supported by the Office of Science, Office of Basic Energy Sciences, of the U.S. Department of Energy under Contract No. DE-AC02--05CH11231. DJ acknowledges support from the Department of Energy Computational Science Graduate Fellowship.


\begin{thebibliography}{89}%
\makeatletter
\providecommand \@ifxundefined [1]{%
 \@ifx{#1\undefined}
}%
\providecommand \@ifnum [1]{%
 \ifnum #1\expandafter \@firstoftwo
 \else \expandafter \@secondoftwo
 \fi
}%
\providecommand \@ifx [1]{%
 \ifx #1\expandafter \@firstoftwo
 \else \expandafter \@secondoftwo
 \fi
}%
\providecommand \natexlab [1]{#1}%
\providecommand \enquote  [1]{``#1''}%
\providecommand \bibnamefont  [1]{#1}%
\providecommand \bibfnamefont [1]{#1}%
\providecommand \citenamefont [1]{#1}%
\providecommand \href@noop [0]{\@secondoftwo}%
\providecommand \href [0]{\begingroup \@sanitize@url \@href}%
\providecommand \@href[1]{\@@startlink{#1}\@@href}%
\providecommand \@@href[1]{\endgroup#1\@@endlink}%
\providecommand \@sanitize@url [0]{\catcode `\\12\catcode `\$12\catcode
  `\&12\catcode `\#12\catcode `\^12\catcode `\_12\catcode `\%12\relax}%
\providecommand \@@startlink[1]{}%
\providecommand \@@endlink[0]{}%
\providecommand \url  [0]{\begingroup\@sanitize@url \@url }%
\providecommand \@url [1]{\endgroup\@href {#1}{\urlprefix }}%
\providecommand \urlprefix  [0]{URL }%
\providecommand \Eprint [0]{\href }%
\providecommand \doibase [0]{http://dx.doi.org/}%
\providecommand \selectlanguage [0]{\@gobble}%
\providecommand \bibinfo  [0]{\@secondoftwo}%
\providecommand \bibfield  [0]{\@secondoftwo}%
\providecommand \translation [1]{[#1]}%
\providecommand \BibitemOpen [0]{}%
\providecommand \bibitemStop [0]{}%
\providecommand \bibitemNoStop [0]{.\EOS\space}%
\providecommand \EOS [0]{\spacefactor3000\relax}%
\providecommand \BibitemShut  [1]{\csname bibitem#1\endcsname}%
\let\auto@bib@innerbib\@empty
\bibitem [{\citenamefont {Gillespie}(2007)}]{gillespie2007stochastic}%
  \BibitemOpen
  \bibfield  {author} {\bibinfo {author} {\bibfnamefont {D.~T.}\ \bibnamefont
  {Gillespie}},\ }\href@noop {} {\bibfield  {journal} {\bibinfo  {journal}
  {Annu. Rev. Phys. Chem.}\ }\textbf {\bibinfo {volume} {58}},\ \bibinfo
  {pages} {35} (\bibinfo {year} {2007})}\BibitemShut {NoStop}%
\bibitem [{\citenamefont {McGrath}\ \emph {et~al.}(2017)\citenamefont
  {McGrath}, \citenamefont {Jones}, \citenamefont {ten Wolde},\ and\
  \citenamefont {Ouldridge}}]{mcgrath2017biochemical}%
  \BibitemOpen
  \bibfield  {author} {\bibinfo {author} {\bibfnamefont {T.}~\bibnamefont
  {McGrath}}, \bibinfo {author} {\bibfnamefont {N.~S.}\ \bibnamefont {Jones}},
  \bibinfo {author} {\bibfnamefont {P.~R.}\ \bibnamefont {ten Wolde}}, \ and\
  \bibinfo {author} {\bibfnamefont {T.~E.}\ \bibnamefont {Ouldridge}},\
  }\href@noop {} {\bibfield  {journal} {\bibinfo  {journal} {Physical Review
  Letters}\ }\textbf {\bibinfo {volume} {118}},\ \bibinfo {pages} {028101}
  (\bibinfo {year} {2017})}\BibitemShut {NoStop}%
\bibitem [{\citenamefont {Seifert}(2012)}]{seifert2012stochastic}%
  \BibitemOpen
  \bibfield  {author} {\bibinfo {author} {\bibfnamefont {U.}~\bibnamefont
  {Seifert}},\ }\href@noop {} {\bibfield  {journal} {\bibinfo  {journal}
  {Reports on progress in Physics}\ }\textbf {\bibinfo {volume} {75}},\
  \bibinfo {pages} {126001} (\bibinfo {year} {2012})}\BibitemShut {NoStop}%
\bibitem [{\citenamefont {Brown}\ and\ \citenamefont
  {Sivak}(2017)}]{brown2017allocating}%
  \BibitemOpen
  \bibfield  {author} {\bibinfo {author} {\bibfnamefont {A.~I.}\ \bibnamefont
  {Brown}}\ and\ \bibinfo {author} {\bibfnamefont {D.~A.}\ \bibnamefont
  {Sivak}},\ }\href@noop {} {\bibfield  {journal} {\bibinfo  {journal}
  {Proceedings of the National Academy of Sciences}\ }\textbf {\bibinfo
  {volume} {114}},\ \bibinfo {pages} {11057} (\bibinfo {year}
  {2017})}\BibitemShut {NoStop}%
\bibitem [{\citenamefont {Visco}\ \emph {et~al.}(2006)\citenamefont {Visco},
  \citenamefont {Puglisi}, \citenamefont {Barrat}, \citenamefont {Trizac},\
  and\ \citenamefont {van Wijland}}]{visco2006fluctuations}%
  \BibitemOpen
  \bibfield  {author} {\bibinfo {author} {\bibfnamefont {P.}~\bibnamefont
  {Visco}}, \bibinfo {author} {\bibfnamefont {A.}~\bibnamefont {Puglisi}},
  \bibinfo {author} {\bibfnamefont {A.}~\bibnamefont {Barrat}}, \bibinfo
  {author} {\bibfnamefont {E.}~\bibnamefont {Trizac}}, \ and\ \bibinfo {author}
  {\bibfnamefont {F.}~\bibnamefont {van Wijland}},\ }\href@noop {} {\bibfield
  {journal} {\bibinfo  {journal} {Journal of statistical Physics}\ }\textbf
  {\bibinfo {volume} {125}},\ \bibinfo {pages} {533} (\bibinfo {year}
  {2006})}\BibitemShut {NoStop}%
\bibitem [{\citenamefont {Chou}\ \emph {et~al.}(2011)\citenamefont {Chou},
  \citenamefont {Mallick},\ and\ \citenamefont {Zia}}]{chou2011non}%
  \BibitemOpen
  \bibfield  {author} {\bibinfo {author} {\bibfnamefont {T.}~\bibnamefont
  {Chou}}, \bibinfo {author} {\bibfnamefont {K.}~\bibnamefont {Mallick}}, \
  and\ \bibinfo {author} {\bibfnamefont {R.}~\bibnamefont {Zia}},\ }\href@noop
  {} {\bibfield  {journal} {\bibinfo  {journal} {Reports on progress in
  Physics}\ }\textbf {\bibinfo {volume} {74}},\ \bibinfo {pages} {116601}
  (\bibinfo {year} {2011})}\BibitemShut {NoStop}%
\bibitem [{\citenamefont {Vaikuntanathan}\ \emph {et~al.}(2014)\citenamefont
  {Vaikuntanathan}, \citenamefont {Gingrich},\ and\ \citenamefont
  {Geissler}}]{vaikuntanathan2014dynamic}%
  \BibitemOpen
  \bibfield  {author} {\bibinfo {author} {\bibfnamefont {S.}~\bibnamefont
  {Vaikuntanathan}}, \bibinfo {author} {\bibfnamefont {T.~R.}\ \bibnamefont
  {Gingrich}}, \ and\ \bibinfo {author} {\bibfnamefont {P.~L.}\ \bibnamefont
  {Geissler}},\ }\href@noop {} {\bibfield  {journal} {\bibinfo  {journal}
  {Physical Review E}\ }\textbf {\bibinfo {volume} {89}},\ \bibinfo {pages}
  {062108} (\bibinfo {year} {2014})}\BibitemShut {NoStop}%
\bibitem [{\citenamefont {Harris}(2015)}]{harris2015fluctuations}%
  \BibitemOpen
  \bibfield  {author} {\bibinfo {author} {\bibfnamefont {R.~J.}\ \bibnamefont
  {Harris}},\ }\href@noop {} {\bibfield  {journal} {\bibinfo  {journal}
  {Journal of Statistical Mechanics: Theory and Experiment}\ }\textbf {\bibinfo
  {volume} {2015}},\ \bibinfo {pages} {P07021} (\bibinfo {year}
  {2015})}\BibitemShut {NoStop}%
\bibitem [{\citenamefont {Berthier}(2014)}]{berthier2014nonequilibrium}%
  \BibitemOpen
  \bibfield  {author} {\bibinfo {author} {\bibfnamefont {L.}~\bibnamefont
  {Berthier}},\ }\href@noop {} {\bibfield  {journal} {\bibinfo  {journal}
  {Physical Review Letters}\ }\textbf {\bibinfo {volume} {112}},\ \bibinfo
  {pages} {220602} (\bibinfo {year} {2014})}\BibitemShut {NoStop}%
\bibitem [{\citenamefont {Garrahan}\ \emph {et~al.}(2007)\citenamefont
  {Garrahan}, \citenamefont {Jack}, \citenamefont {Lecomte}, \citenamefont
  {Pitard}, \citenamefont {van Duijvendijk},\ and\ \citenamefont {van
  Wijland}}]{garrahan2007dynamical}%
  \BibitemOpen
  \bibfield  {author} {\bibinfo {author} {\bibfnamefont {J.~P.}\ \bibnamefont
  {Garrahan}}, \bibinfo {author} {\bibfnamefont {R.~L.}\ \bibnamefont {Jack}},
  \bibinfo {author} {\bibfnamefont {V.}~\bibnamefont {Lecomte}}, \bibinfo
  {author} {\bibfnamefont {E.}~\bibnamefont {Pitard}}, \bibinfo {author}
  {\bibfnamefont {K.}~\bibnamefont {van Duijvendijk}}, \ and\ \bibinfo {author}
  {\bibfnamefont {F.}~\bibnamefont {van Wijland}},\ }\href@noop {} {\bibfield
  {journal} {\bibinfo  {journal} {Physical Review Letters}\ }\textbf {\bibinfo
  {volume} {98}},\ \bibinfo {pages} {195702} (\bibinfo {year}
  {2007})}\BibitemShut {NoStop}%
\bibitem [{\citenamefont {Garrahan}\ \emph {et~al.}(2009)\citenamefont
  {Garrahan}, \citenamefont {Jack}, \citenamefont {Lecomte}, \citenamefont
  {Pitard}, \citenamefont {van Duijvendijk},\ and\ \citenamefont {van
  Wijland}}]{garrahan2009first}%
  \BibitemOpen
  \bibfield  {author} {\bibinfo {author} {\bibfnamefont {J.~P.}\ \bibnamefont
  {Garrahan}}, \bibinfo {author} {\bibfnamefont {R.~L.}\ \bibnamefont {Jack}},
  \bibinfo {author} {\bibfnamefont {V.}~\bibnamefont {Lecomte}}, \bibinfo
  {author} {\bibfnamefont {E.}~\bibnamefont {Pitard}}, \bibinfo {author}
  {\bibfnamefont {K.}~\bibnamefont {van Duijvendijk}}, \ and\ \bibinfo {author}
  {\bibfnamefont {F.}~\bibnamefont {van Wijland}},\ }\href@noop {} {\bibfield
  {journal} {\bibinfo  {journal} {Journal of Physics A: Mathematical and
  Theoretical}\ }\textbf {\bibinfo {volume} {42}},\ \bibinfo {pages} {075007}
  (\bibinfo {year} {2009})}\BibitemShut {NoStop}%
\bibitem [{\citenamefont {Ritort}(2008)}]{ritort2008nonequilibrium}%
  \BibitemOpen
  \bibfield  {author} {\bibinfo {author} {\bibfnamefont {F.}~\bibnamefont
  {Ritort}},\ }\href@noop {} {\bibfield  {journal} {\bibinfo  {journal}
  {Advances in Chemical Physics}\ }\textbf {\bibinfo {volume} {137}},\ \bibinfo
  {pages} {31} (\bibinfo {year} {2008})}\BibitemShut {NoStop}%
\bibitem [{\citenamefont {Seifert}(2005)}]{seifert2005entropy}%
  \BibitemOpen
  \bibfield  {author} {\bibinfo {author} {\bibfnamefont {U.}~\bibnamefont
  {Seifert}},\ }\href@noop {} {\bibfield  {journal} {\bibinfo  {journal}
  {Physical Review Letters}\ }\textbf {\bibinfo {volume} {95}},\ \bibinfo
  {pages} {040602} (\bibinfo {year} {2005})}\BibitemShut {NoStop}%
\bibitem [{\citenamefont {Speck}\ \emph {et~al.}(2012)\citenamefont {Speck},
  \citenamefont {Engel},\ and\ \citenamefont {Seifert}}]{speck2012large}%
  \BibitemOpen
  \bibfield  {author} {\bibinfo {author} {\bibfnamefont {T.}~\bibnamefont
  {Speck}}, \bibinfo {author} {\bibfnamefont {A.}~\bibnamefont {Engel}}, \ and\
  \bibinfo {author} {\bibfnamefont {U.}~\bibnamefont {Seifert}},\ }\href@noop
  {} {\bibfield  {journal} {\bibinfo  {journal} {Journal of Statistical
  Mechanics: Theory and Experiment}\ }\textbf {\bibinfo {volume} {2012}},\
  \bibinfo {pages} {P12001} (\bibinfo {year} {2012})}\BibitemShut {NoStop}%
\bibitem [{\citenamefont {Lecomte}\ \emph {et~al.}(2010)\citenamefont
  {Lecomte}, \citenamefont {Imparato},\ and\ \citenamefont
  {Wijland}}]{lecomte2010current}%
  \BibitemOpen
  \bibfield  {author} {\bibinfo {author} {\bibfnamefont {V.}~\bibnamefont
  {Lecomte}}, \bibinfo {author} {\bibfnamefont {A.}~\bibnamefont {Imparato}}, \
  and\ \bibinfo {author} {\bibfnamefont {F.~v.}\ \bibnamefont {Wijland}},\
  }\href@noop {} {\bibfield  {journal} {\bibinfo  {journal} {Progress of
  Theoretical Physics Supplement}\ }\textbf {\bibinfo {volume} {184}},\
  \bibinfo {pages} {276} (\bibinfo {year} {2010})}\BibitemShut {NoStop}%
\bibitem [{\citenamefont {Gingrich}\ and\ \citenamefont
  {Horowitz}(2017)}]{gingrich2017fundamental}%
  \BibitemOpen
  \bibfield  {author} {\bibinfo {author} {\bibfnamefont {T.~R.}\ \bibnamefont
  {Gingrich}}\ and\ \bibinfo {author} {\bibfnamefont {J.~M.}\ \bibnamefont
  {Horowitz}},\ }\href@noop {} {\bibfield  {journal} {\bibinfo  {journal}
  {Physical Review Letters}\ }\textbf {\bibinfo {volume} {119}},\ \bibinfo
  {pages} {170601} (\bibinfo {year} {2017})}\BibitemShut {NoStop}%
\bibitem [{\citenamefont {Fodor}\ \emph {et~al.}(2015)\citenamefont {Fodor},
  \citenamefont {Guo}, \citenamefont {Gov}, \citenamefont {Visco},
  \citenamefont {Weitz},\ and\ \citenamefont {van
  Wijland}}]{fodor2015activity}%
  \BibitemOpen
  \bibfield  {author} {\bibinfo {author} {\bibfnamefont {{\'E}.}~\bibnamefont
  {Fodor}}, \bibinfo {author} {\bibfnamefont {M.}~\bibnamefont {Guo}}, \bibinfo
  {author} {\bibfnamefont {N.}~\bibnamefont {Gov}}, \bibinfo {author}
  {\bibfnamefont {P.}~\bibnamefont {Visco}}, \bibinfo {author} {\bibfnamefont
  {D.}~\bibnamefont {Weitz}}, \ and\ \bibinfo {author} {\bibfnamefont
  {F.}~\bibnamefont {van Wijland}},\ }\href@noop {} {\bibfield  {journal}
  {\bibinfo  {journal} {EPL (EuroPhysics Letters)}\ }\textbf {\bibinfo {volume}
  {110}},\ \bibinfo {pages} {48005} (\bibinfo {year} {2015})}\BibitemShut
  {NoStop}%
\bibitem [{\citenamefont {Chandler}(1987)}]{chandler1987introduction}%
  \BibitemOpen
  \bibfield  {author} {\bibinfo {author} {\bibfnamefont {D.}~\bibnamefont
  {Chandler}},\ }\href@noop {} {\bibfield  {journal} {\bibinfo  {journal}
  {Introduction to Modern Statistical Mechanics, by David Chandler, pp. 288.
  Foreword by David Chandler. Oxford University Press, Sep 1987. ISBN-10:
  0195042778. ISBN-13: 9780195042771}\ ,\ \bibinfo {pages} {288}} (\bibinfo
  {year} {1987})}\BibitemShut {NoStop}%
\bibitem [{\citenamefont {Binney}\ \emph {et~al.}(1992)\citenamefont {Binney},
  \citenamefont {Dowrick}, \citenamefont {Fisher},\ and\ \citenamefont
  {Newman}}]{binney1992theory}%
  \BibitemOpen
  \bibfield  {author} {\bibinfo {author} {\bibfnamefont {J.~J.}\ \bibnamefont
  {Binney}}, \bibinfo {author} {\bibfnamefont {N.}~\bibnamefont {Dowrick}},
  \bibinfo {author} {\bibfnamefont {A.}~\bibnamefont {Fisher}}, \ and\ \bibinfo
  {author} {\bibfnamefont {M.}~\bibnamefont {Newman}},\ }\href@noop {} {\emph
  {\bibinfo {title} {The theory of critical phenomena: an introduction to the
  renormalization group}}}\ (\bibinfo  {publisher} {Oxford University Press,
  Inc.},\ \bibinfo {year} {1992})\BibitemShut {NoStop}%
\bibitem [{\citenamefont {Gallavotti}\ and\ \citenamefont
  {Cohen}(1995)}]{gallavotti1995dynamical}%
  \BibitemOpen
  \bibfield  {author} {\bibinfo {author} {\bibfnamefont {G.}~\bibnamefont
  {Gallavotti}}\ and\ \bibinfo {author} {\bibfnamefont {E.~G.~D.}\ \bibnamefont
  {Cohen}},\ }\href@noop {} {\bibfield  {journal} {\bibinfo  {journal}
  {Physical Review Letters}\ }\textbf {\bibinfo {volume} {74}},\ \bibinfo
  {pages} {2694} (\bibinfo {year} {1995})}\BibitemShut {NoStop}%
\bibitem [{\citenamefont {Maes}(1999)}]{maes1999fluctuation}%
  \BibitemOpen
  \bibfield  {author} {\bibinfo {author} {\bibfnamefont {C.}~\bibnamefont
  {Maes}},\ }\href@noop {} {\bibfield  {journal} {\bibinfo  {journal} {Journal
  of Statistical Physics}\ }\textbf {\bibinfo {volume} {95}},\ \bibinfo {pages}
  {367} (\bibinfo {year} {1999})}\BibitemShut {NoStop}%
\bibitem [{\citenamefont {Jarzynski}(1997)}]{jarzynski1997nonequilibrium}%
  \BibitemOpen
  \bibfield  {author} {\bibinfo {author} {\bibfnamefont {C.}~\bibnamefont
  {Jarzynski}},\ }\href@noop {} {\bibfield  {journal} {\bibinfo  {journal}
  {Physical Review Letters}\ }\textbf {\bibinfo {volume} {78}},\ \bibinfo
  {pages} {2690} (\bibinfo {year} {1997})}\BibitemShut {NoStop}%
\bibitem [{\citenamefont {Kurchan}(1998)}]{kurchan1998fluctuation}%
  \BibitemOpen
  \bibfield  {author} {\bibinfo {author} {\bibfnamefont {J.}~\bibnamefont
  {Kurchan}},\ }\href@noop {} {\bibfield  {journal} {\bibinfo  {journal}
  {Journal of Physics A: Mathematical and General}\ }\textbf {\bibinfo {volume}
  {31}},\ \bibinfo {pages} {3719} (\bibinfo {year} {1998})}\BibitemShut
  {NoStop}%
\bibitem [{\citenamefont {Lebowitz}\ and\ \citenamefont
  {Spohn}(1999)}]{lebowitz1999gallavotti}%
  \BibitemOpen
  \bibfield  {author} {\bibinfo {author} {\bibfnamefont {J.~L.}\ \bibnamefont
  {Lebowitz}}\ and\ \bibinfo {author} {\bibfnamefont {H.}~\bibnamefont
  {Spohn}},\ }\href@noop {} {\bibfield  {journal} {\bibinfo  {journal} {Journal
  of Statistical Physics}\ }\textbf {\bibinfo {volume} {95}},\ \bibinfo {pages}
  {333} (\bibinfo {year} {1999})}\BibitemShut {NoStop}%
\bibitem [{\citenamefont {Crooks}(1999)}]{crooks1999entropy}%
  \BibitemOpen
  \bibfield  {author} {\bibinfo {author} {\bibfnamefont {G.~E.}\ \bibnamefont
  {Crooks}},\ }\href@noop {} {\bibfield  {journal} {\bibinfo  {journal}
  {Physical Review E}\ }\textbf {\bibinfo {volume} {60}},\ \bibinfo {pages}
  {2721} (\bibinfo {year} {1999})}\BibitemShut {NoStop}%
\bibitem [{\citenamefont {Harris}\ and\ \citenamefont
  {Sch{\"u}tz}(2007)}]{harris2007fluctuation}%
  \BibitemOpen
  \bibfield  {author} {\bibinfo {author} {\bibfnamefont {R.~J.}\ \bibnamefont
  {Harris}}\ and\ \bibinfo {author} {\bibfnamefont {G.~M.}\ \bibnamefont
  {Sch{\"u}tz}},\ }\href@noop {} {\bibfield  {journal} {\bibinfo  {journal}
  {Journal of Statistical Mechanics: Theory and Experiment}\ }\textbf {\bibinfo
  {volume} {2007}},\ \bibinfo {pages} {P07020} (\bibinfo {year}
  {2007})}\BibitemShut {NoStop}%
\bibitem [{\citenamefont {Den~Hollander}(2008)}]{den2008large}%
  \BibitemOpen
  \bibfield  {author} {\bibinfo {author} {\bibfnamefont {F.}~\bibnamefont
  {Den~Hollander}},\ }\href@noop {} {\emph {\bibinfo {title} {Large
  Deviations}}},\ Vol.~\bibinfo {volume} {14}\ (\bibinfo  {publisher} {American
  Mathematical Soc.},\ \bibinfo {year} {2008})\BibitemShut {NoStop}%
\bibitem [{\citenamefont {Touchette}(2009)}]{touchette2009large}%
  \BibitemOpen
  \bibfield  {author} {\bibinfo {author} {\bibfnamefont {H.}~\bibnamefont
  {Touchette}},\ }\href@noop {} {\bibfield  {journal} {\bibinfo  {journal}
  {Physics Reports}\ }\textbf {\bibinfo {volume} {478}},\ \bibinfo {pages} {1}
  (\bibinfo {year} {2009})}\BibitemShut {NoStop}%
\bibitem [{Note1()}]{Note1}%
  \BibitemOpen
  \bibinfo {note} {More directly it quantifies the rate of decay of a
  fluctuation $a$, which depends both on the likelihood of $a$ {\protect \em
  and} the basic timescale governing the establishment and decay of
  fluctuations. This latter piece plays a key role for certain models~\cite
  {spiliopoulos2013large,bouchet2016large,whitelam2018large}.}\BibitemShut
  {Stop}%
\bibitem [{\citenamefont {Giardina}\ \emph {et~al.}(2006)\citenamefont
  {Giardina}, \citenamefont {Kurchan},\ and\ \citenamefont
  {Peliti}}]{giardina2006direct}%
  \BibitemOpen
  \bibfield  {author} {\bibinfo {author} {\bibfnamefont {C.}~\bibnamefont
  {Giardina}}, \bibinfo {author} {\bibfnamefont {J.}~\bibnamefont {Kurchan}}, \
  and\ \bibinfo {author} {\bibfnamefont {L.}~\bibnamefont {Peliti}},\
  }\href@noop {} {\bibfield  {journal} {\bibinfo  {journal} {Physical Review
  Letters}\ }\textbf {\bibinfo {volume} {96}},\ \bibinfo {pages} {120603}
  (\bibinfo {year} {2006})}\BibitemShut {NoStop}%
\bibitem [{\citenamefont {Chetrite}\ and\ \citenamefont
  {Touchette}(2015)}]{chetrite2015variational}%
  \BibitemOpen
  \bibfield  {author} {\bibinfo {author} {\bibfnamefont {R.}~\bibnamefont
  {Chetrite}}\ and\ \bibinfo {author} {\bibfnamefont {H.}~\bibnamefont
  {Touchette}},\ }\href@noop {} {\bibfield  {journal} {\bibinfo  {journal}
  {Journal of Statistical Mechanics: Theory and Experiment}\ }\textbf {\bibinfo
  {volume} {2015}},\ \bibinfo {pages} {P12001} (\bibinfo {year}
  {2015})}\BibitemShut {NoStop}%
\bibitem [{\citenamefont {Jack}\ and\ \citenamefont
  {Sollich}(2015)}]{jack2015effective}%
  \BibitemOpen
  \bibfield  {author} {\bibinfo {author} {\bibfnamefont {R.~L.}\ \bibnamefont
  {Jack}}\ and\ \bibinfo {author} {\bibfnamefont {P.}~\bibnamefont {Sollich}},\
  }\href@noop {} {\bibfield  {journal} {\bibinfo  {journal} {The European
  Physical Journal Special Topics}\ }\textbf {\bibinfo {volume} {224}},\
  \bibinfo {pages} {2351} (\bibinfo {year} {2015})}\BibitemShut {NoStop}%
\bibitem [{\citenamefont {Ray}\ \emph {et~al.}(2018)\citenamefont {Ray},
  \citenamefont {Chan},\ and\ \citenamefont {Limmer}}]{ray2018exact}%
  \BibitemOpen
  \bibfield  {author} {\bibinfo {author} {\bibfnamefont {U.}~\bibnamefont
  {Ray}}, \bibinfo {author} {\bibfnamefont {G.~K.-L.}\ \bibnamefont {Chan}}, \
  and\ \bibinfo {author} {\bibfnamefont {D.~T.}\ \bibnamefont {Limmer}},\
  }\href@noop {} {\bibfield  {journal} {\bibinfo  {journal} {Physical Review
  Letters}\ }\textbf {\bibinfo {volume} {120}},\ \bibinfo {pages} {210602}
  (\bibinfo {year} {2018})}\BibitemShut {NoStop}%
\bibitem [{\citenamefont {Nemoto}\ \emph {et~al.}(2017)\citenamefont {Nemoto},
  \citenamefont {Jack},\ and\ \citenamefont {Lecomte}}]{nemoto2017finite}%
  \BibitemOpen
  \bibfield  {author} {\bibinfo {author} {\bibfnamefont {T.}~\bibnamefont
  {Nemoto}}, \bibinfo {author} {\bibfnamefont {R.~L.}\ \bibnamefont {Jack}}, \
  and\ \bibinfo {author} {\bibfnamefont {V.}~\bibnamefont {Lecomte}},\
  }\href@noop {} {\bibfield  {journal} {\bibinfo  {journal} {Physical Review
  Letters}\ }\textbf {\bibinfo {volume} {118}},\ \bibinfo {pages} {115702}
  (\bibinfo {year} {2017})}\BibitemShut {NoStop}%
\bibitem [{\citenamefont {Ferr{\'e}}\ and\ \citenamefont
  {Touchette}(2018)}]{ferre2018adaptive}%
  \BibitemOpen
  \bibfield  {author} {\bibinfo {author} {\bibfnamefont {G.}~\bibnamefont
  {Ferr{\'e}}}\ and\ \bibinfo {author} {\bibfnamefont {H.}~\bibnamefont
  {Touchette}},\ }\href@noop {} {\bibfield  {journal} {\bibinfo  {journal}
  {arXiv preprint arXiv:1803.11117}\ } (\bibinfo {year} {2018})}\BibitemShut
  {NoStop}%
\bibitem [{\citenamefont {Chetrite}\ and\ \citenamefont
  {Touchette}(2013)}]{chetrite2013nonequilibrium}%
  \BibitemOpen
  \bibfield  {author} {\bibinfo {author} {\bibfnamefont {R.}~\bibnamefont
  {Chetrite}}\ and\ \bibinfo {author} {\bibfnamefont {H.}~\bibnamefont
  {Touchette}},\ }\href@noop {} {\bibfield  {journal} {\bibinfo  {journal}
  {Physical Review Letters}\ }\textbf {\bibinfo {volume} {111}},\ \bibinfo
  {pages} {120601} (\bibinfo {year} {2013})}\BibitemShut {NoStop}%
\bibitem [{\citenamefont {Lecomte}\ and\ \citenamefont
  {Tailleur}(2007)}]{lecomte2007numerical}%
  \BibitemOpen
  \bibfield  {author} {\bibinfo {author} {\bibfnamefont {V.}~\bibnamefont
  {Lecomte}}\ and\ \bibinfo {author} {\bibfnamefont {J.}~\bibnamefont
  {Tailleur}},\ }\href@noop {} {\bibfield  {journal} {\bibinfo  {journal}
  {Journal of Statistical Mechanics: Theory and Experiment}\ }\textbf {\bibinfo
  {volume} {2007}},\ \bibinfo {pages} {P03004} (\bibinfo {year}
  {2007})}\BibitemShut {NoStop}%
\bibitem [{\citenamefont {Kundu}\ \emph {et~al.}(2011)\citenamefont {Kundu},
  \citenamefont {Sabhapandit},\ and\ \citenamefont
  {Dhar}}]{kundu2011application}%
  \BibitemOpen
  \bibfield  {author} {\bibinfo {author} {\bibfnamefont {A.}~\bibnamefont
  {Kundu}}, \bibinfo {author} {\bibfnamefont {S.}~\bibnamefont {Sabhapandit}},
  \ and\ \bibinfo {author} {\bibfnamefont {A.}~\bibnamefont {Dhar}},\
  }\href@noop {} {\bibfield  {journal} {\bibinfo  {journal} {Physical Review
  E}\ }\textbf {\bibinfo {volume} {83}},\ \bibinfo {pages} {031119} (\bibinfo
  {year} {2011})}\BibitemShut {NoStop}%
\bibitem [{\citenamefont {Klymko}\ \emph {et~al.}(2018)\citenamefont {Klymko},
  \citenamefont {Geissler}, \citenamefont {Garrahan},\ and\ \citenamefont
  {Whitelam}}]{klymko2017rare}%
  \BibitemOpen
  \bibfield  {author} {\bibinfo {author} {\bibfnamefont {K.}~\bibnamefont
  {Klymko}}, \bibinfo {author} {\bibfnamefont {P.~L.}\ \bibnamefont
  {Geissler}}, \bibinfo {author} {\bibfnamefont {J.~P.}\ \bibnamefont
  {Garrahan}}, \ and\ \bibinfo {author} {\bibfnamefont {S.}~\bibnamefont
  {Whitelam}},\ }\href {\doibase 10.1103/PhysRevE.97.032123} {\bibfield
  {journal} {\bibinfo  {journal} {Phys. Rev. E}\ }\textbf {\bibinfo {volume}
  {97}},\ \bibinfo {pages} {032123} (\bibinfo {year} {2018})}\BibitemShut
  {NoStop}%
\bibitem [{\citenamefont
  {Whitelam}(2018{\natexlab{a}})}]{whitelam2018sampling}%
  \BibitemOpen
  \bibfield  {author} {\bibinfo {author} {\bibfnamefont {S.}~\bibnamefont
  {Whitelam}},\ }\href@noop {} {\bibfield  {journal} {\bibinfo  {journal}
  {Physical Review E}\ }\textbf {\bibinfo {volume} {97}},\ \bibinfo {pages}
  {032122} (\bibinfo {year} {2018}{\natexlab{a}})}\BibitemShut {NoStop}%
\bibitem [{\citenamefont {Whitelam}(2018{\natexlab{b}})}]{whitelam2018multi}%
  \BibitemOpen
  \bibfield  {author} {\bibinfo {author} {\bibfnamefont {S.}~\bibnamefont
  {Whitelam}},\ }\href {http://stacks.iop.org/1742-5468/2018/i=6/a=063211}
  {\bibfield  {journal} {\bibinfo  {journal} {Journal of Statistical Mechanics:
  Theory and Experiment}\ }\textbf {\bibinfo {volume} {2018}},\ \bibinfo
  {pages} {063211} (\bibinfo {year} {2018}{\natexlab{b}})}\BibitemShut
  {NoStop}%
\bibitem [{\citenamefont {Bucklew}\ \emph {et~al.}(1990)\citenamefont
  {Bucklew}, \citenamefont {Ney},\ and\ \citenamefont
  {Sadowsky}}]{bucklew1990monte}%
  \BibitemOpen
  \bibfield  {author} {\bibinfo {author} {\bibfnamefont {J.~A.}\ \bibnamefont
  {Bucklew}}, \bibinfo {author} {\bibfnamefont {P.}~\bibnamefont {Ney}}, \ and\
  \bibinfo {author} {\bibfnamefont {J.~S.}\ \bibnamefont {Sadowsky}},\
  }\href@noop {} {\bibfield  {journal} {\bibinfo  {journal} {Journal of Applied
  Probability}\ }\textbf {\bibinfo {volume} {27}},\ \bibinfo {pages} {44}
  (\bibinfo {year} {1990})}\BibitemShut {NoStop}%
\bibitem [{\citenamefont {Glynn}\ and\ \citenamefont
  {Iglehart}(1989)}]{glynn1989importance}%
  \BibitemOpen
  \bibfield  {author} {\bibinfo {author} {\bibfnamefont {P.~W.}\ \bibnamefont
  {Glynn}}\ and\ \bibinfo {author} {\bibfnamefont {D.~L.}\ \bibnamefont
  {Iglehart}},\ }\href@noop {} {\bibfield  {journal} {\bibinfo  {journal}
  {Management Science}\ }\textbf {\bibinfo {volume} {35}},\ \bibinfo {pages}
  {1367} (\bibinfo {year} {1989})}\BibitemShut {NoStop}%
\bibitem [{\citenamefont {Sadowsky}\ and\ \citenamefont
  {Bucklew}(1990)}]{sadowsky1990large}%
  \BibitemOpen
  \bibfield  {author} {\bibinfo {author} {\bibfnamefont {J.~S.}\ \bibnamefont
  {Sadowsky}}\ and\ \bibinfo {author} {\bibfnamefont {J.~A.}\ \bibnamefont
  {Bucklew}},\ }\href@noop {} {\bibfield  {journal} {\bibinfo  {journal} {IEEE
  transactions on Information Theory}\ }\textbf {\bibinfo {volume} {36}},\
  \bibinfo {pages} {579} (\bibinfo {year} {1990})}\BibitemShut {NoStop}%
\bibitem [{\citenamefont {Glasserman}\ \emph {et~al.}(1997)\citenamefont
  {Glasserman}, \citenamefont {Wang} \emph
  {et~al.}}]{glasserman1997counterexamples}%
  \BibitemOpen
  \bibfield  {author} {\bibinfo {author} {\bibfnamefont {P.}~\bibnamefont
  {Glasserman}}, \bibinfo {author} {\bibfnamefont {Y.}~\bibnamefont {Wang}},
  \emph {et~al.},\ }\href@noop {} {\bibfield  {journal} {\bibinfo  {journal}
  {The Annals of Applied Probability}\ }\textbf {\bibinfo {volume} {7}},\
  \bibinfo {pages} {731} (\bibinfo {year} {1997})}\BibitemShut {NoStop}%
\bibitem [{\citenamefont {Warren}\ and\ \citenamefont
  {Allen}(2018)}]{warren2018trajectory}%
  \BibitemOpen
  \bibfield  {author} {\bibinfo {author} {\bibfnamefont {P.~B.}\ \bibnamefont
  {Warren}}\ and\ \bibinfo {author} {\bibfnamefont {R.~J.}\ \bibnamefont
  {Allen}},\ }\href@noop {} {\bibfield  {journal} {\bibinfo  {journal}
  {Molecular Physics}\ }\textbf {\bibinfo {volume} {116}},\ \bibinfo {pages}
  {3104} (\bibinfo {year} {2018})}\BibitemShut {NoStop}%
\bibitem [{\citenamefont {Bucklew}(1990)}]{bucklew1990large}%
  \BibitemOpen
  \bibfield  {author} {\bibinfo {author} {\bibfnamefont {J.~A.}\ \bibnamefont
  {Bucklew}},\ }\href@noop {} {\emph {\bibinfo {title} {Large deviation
  techniques in decision, simulation, and estimation}}}\ (\bibinfo  {publisher}
  {Wiley New York},\ \bibinfo {year} {1990})\BibitemShut {NoStop}%
\bibitem [{\citenamefont {Asmussen}\ and\ \citenamefont
  {Glynn}(2007)}]{asmussen2007stochastic}%
  \BibitemOpen
  \bibfield  {author} {\bibinfo {author} {\bibfnamefont {S.}~\bibnamefont
  {Asmussen}}\ and\ \bibinfo {author} {\bibfnamefont {P.~W.}\ \bibnamefont
  {Glynn}},\ }\href@noop {} {\emph {\bibinfo {title} {Stochastic simulation:
  algorithms and analysis}}},\ Vol.~\bibinfo {volume} {57}\ (\bibinfo
  {publisher} {Springer Science \& Business Media},\ \bibinfo {year}
  {2007})\BibitemShut {NoStop}%
\bibitem [{\citenamefont {Juneja}\ and\ \citenamefont
  {Shahabuddin}(2006)}]{juneja2006rare}%
  \BibitemOpen
  \bibfield  {author} {\bibinfo {author} {\bibfnamefont {S.}~\bibnamefont
  {Juneja}}\ and\ \bibinfo {author} {\bibfnamefont {P.}~\bibnamefont
  {Shahabuddin}},\ }\href@noop {} {\bibfield  {journal} {\bibinfo  {journal}
  {Handbooks in operations research and management science}\ }\textbf {\bibinfo
  {volume} {13}},\ \bibinfo {pages} {291} (\bibinfo {year} {2006})}\BibitemShut
  {NoStop}%
\bibitem [{\citenamefont {Bucklew}(2013)}]{bucklew2013introduction}%
  \BibitemOpen
  \bibfield  {author} {\bibinfo {author} {\bibfnamefont {J.}~\bibnamefont
  {Bucklew}},\ }\href@noop {} {\emph {\bibinfo {title} {Introduction to rare
  event simulation}}}\ (\bibinfo  {publisher} {Springer Science \& Business
  Media},\ \bibinfo {year} {2013})\BibitemShut {NoStop}%
\bibitem [{Note2()}]{Note2}%
  \BibitemOpen
  \bibinfo {note} {The ``typical'' value of a time-integrated observable is the
  value to which the sample mean of an unbiased estimator of that quantity
  converges at long times, provided that a large-deviation principle exists and
  that the associated rate function has a unique zero~\cite
  {touchette2009large}.}\BibitemShut {Stop}%
\bibitem [{\citenamefont {Varadhan}(2010)}]{varadhan2010large}%
  \BibitemOpen
  \bibfield  {author} {\bibinfo {author} {\bibfnamefont {S.~S.}\ \bibnamefont
  {Varadhan}},\ }in\ \href@noop {} {\emph {\bibinfo {booktitle} {Proceedings of
  the International Congress of Mathematicians 2010 (ICM 2010) (In 4 Volumes)
  Vol. I: Plenary Lectures and Ceremonies Vols. II--IV: Invited Lectures}}}\
  (\bibinfo {organization} {World Scientific},\ \bibinfo {year} {2010})\ pp.\
  \bibinfo {pages} {622--639}\BibitemShut {NoStop}%
\bibitem [{\citenamefont {Pietzonka}\ \emph {et~al.}(2016)\citenamefont
  {Pietzonka}, \citenamefont {Barato},\ and\ \citenamefont
  {Seifert}}]{pietzonka2016universal}%
  \BibitemOpen
  \bibfield  {author} {\bibinfo {author} {\bibfnamefont {P.}~\bibnamefont
  {Pietzonka}}, \bibinfo {author} {\bibfnamefont {A.~C.}\ \bibnamefont
  {Barato}}, \ and\ \bibinfo {author} {\bibfnamefont {U.}~\bibnamefont
  {Seifert}},\ }\href@noop {} {\bibfield  {journal} {\bibinfo  {journal}
  {Physical Review E}\ }\textbf {\bibinfo {volume} {93}},\ \bibinfo {pages}
  {052145} (\bibinfo {year} {2016})}\BibitemShut {NoStop}%
\bibitem [{\citenamefont {Gingrich}\ \emph {et~al.}(2016)\citenamefont
  {Gingrich}, \citenamefont {Horowitz}, \citenamefont {Perunov},\ and\
  \citenamefont {England}}]{gingrich2016dissipation}%
  \BibitemOpen
  \bibfield  {author} {\bibinfo {author} {\bibfnamefont {T.~R.}\ \bibnamefont
  {Gingrich}}, \bibinfo {author} {\bibfnamefont {J.~M.}\ \bibnamefont
  {Horowitz}}, \bibinfo {author} {\bibfnamefont {N.}~\bibnamefont {Perunov}}, \
  and\ \bibinfo {author} {\bibfnamefont {J.~L.}\ \bibnamefont {England}},\
  }\href@noop {} {\bibfield  {journal} {\bibinfo  {journal} {Physical Review
  Letters}\ }\textbf {\bibinfo {volume} {116}},\ \bibinfo {pages} {120601}
  (\bibinfo {year} {2016})}\BibitemShut {NoStop}%
\bibitem [{\citenamefont {Garrahan}(2017)}]{garrahan2017simple}%
  \BibitemOpen
  \bibfield  {author} {\bibinfo {author} {\bibfnamefont {J.~P.}\ \bibnamefont
  {Garrahan}},\ }\href@noop {} {\bibfield  {journal} {\bibinfo  {journal}
  {Physical Review E}\ }\textbf {\bibinfo {volume} {95}},\ \bibinfo {pages}
  {032134} (\bibinfo {year} {2017})}\BibitemShut {NoStop}%
\bibitem [{\citenamefont {Maes}\ and\ \citenamefont
  {Neto{\v{c}}n{\`y}}(2008)}]{maes2008canonical}%
  \BibitemOpen
  \bibfield  {author} {\bibinfo {author} {\bibfnamefont {C.}~\bibnamefont
  {Maes}}\ and\ \bibinfo {author} {\bibfnamefont {K.}~\bibnamefont
  {Neto{\v{c}}n{\`y}}},\ }\href@noop {} {\bibfield  {journal} {\bibinfo
  {journal} {EPL (EuroPhysics Letters)}\ }\textbf {\bibinfo {volume} {82}},\
  \bibinfo {pages} {30003} (\bibinfo {year} {2008})}\BibitemShut {NoStop}%
\bibitem [{\citenamefont {Bertini}\ \emph {et~al.}(2015)\citenamefont
  {Bertini}, \citenamefont {Faggionato}, \citenamefont {Gabrielli} \emph
  {et~al.}}]{bertini2015large}%
  \BibitemOpen
  \bibfield  {author} {\bibinfo {author} {\bibfnamefont {L.}~\bibnamefont
  {Bertini}}, \bibinfo {author} {\bibfnamefont {A.}~\bibnamefont {Faggionato}},
  \bibinfo {author} {\bibfnamefont {D.}~\bibnamefont {Gabrielli}},  \emph
  {et~al.},\ }in\ \href@noop {} {\emph {\bibinfo {booktitle} {Annales de
  l'Institut Henri Poincar{\'e}, Probabilit{\'e}s et Statistiques}}},\
  Vol.~\bibinfo {volume} {51}\ (\bibinfo {organization} {Institut Henri
  Poincar{\'e}},\ \bibinfo {year} {2015})\ pp.\ \bibinfo {pages}
  {867--900}\BibitemShut {NoStop}%
\bibitem [{\citenamefont {Tsobgni~Nyawo}\ and\ \citenamefont
  {Touchette}(2016)}]{PhysRevE.94.032101}%
  \BibitemOpen
  \bibfield  {author} {\bibinfo {author} {\bibfnamefont {P.}~\bibnamefont
  {Tsobgni~Nyawo}}\ and\ \bibinfo {author} {\bibfnamefont {H.}~\bibnamefont
  {Touchette}},\ }\href {\doibase 10.1103/PhysRevE.94.032101} {\bibfield
  {journal} {\bibinfo  {journal} {Phys. Rev. E}\ }\textbf {\bibinfo {volume}
  {94}},\ \bibinfo {pages} {032101} (\bibinfo {year} {2016})}\BibitemShut
  {NoStop}%
\bibitem [{\citenamefont {Jack}\ and\ \citenamefont
  {Sollich}(2013)}]{jack2013large}%
  \BibitemOpen
  \bibfield  {author} {\bibinfo {author} {\bibfnamefont {R.~L.}\ \bibnamefont
  {Jack}}\ and\ \bibinfo {author} {\bibfnamefont {P.}~\bibnamefont {Sollich}},\
  }\href@noop {} {\bibfield  {journal} {\bibinfo  {journal} {Journal of Physics
  A: Mathematical and Theoretical}\ }\textbf {\bibinfo {volume} {47}},\
  \bibinfo {pages} {015003} (\bibinfo {year} {2013})}\BibitemShut {NoStop}%
\bibitem [{\citenamefont {Ray}\ \emph {et~al.}(2017)\citenamefont {Ray},
  \citenamefont {Chan},\ and\ \citenamefont {Limmer}}]{ray2017importance}%
  \BibitemOpen
  \bibfield  {author} {\bibinfo {author} {\bibfnamefont {U.}~\bibnamefont
  {Ray}}, \bibinfo {author} {\bibfnamefont {G.~K.}\ \bibnamefont {Chan}}, \
  and\ \bibinfo {author} {\bibfnamefont {D.~T.}\ \bibnamefont {Limmer}},\
  }\href@noop {} {\bibfield  {journal} {\bibinfo  {journal} {arXiv preprint
  arXiv:1708.00459}\ } (\bibinfo {year} {2017})}\BibitemShut {NoStop}%
\bibitem [{\citenamefont {Rohwer}\ \emph {et~al.}(2015)\citenamefont {Rohwer},
  \citenamefont {Angeletti},\ and\ \citenamefont
  {Touchette}}]{rohwer2015convergence}%
  \BibitemOpen
  \bibfield  {author} {\bibinfo {author} {\bibfnamefont {C.~M.}\ \bibnamefont
  {Rohwer}}, \bibinfo {author} {\bibfnamefont {F.}~\bibnamefont {Angeletti}}, \
  and\ \bibinfo {author} {\bibfnamefont {H.}~\bibnamefont {Touchette}},\
  }\href@noop {} {\bibfield  {journal} {\bibinfo  {journal} {Physical Review
  E}\ }\textbf {\bibinfo {volume} {92}},\ \bibinfo {pages} {052104} (\bibinfo
  {year} {2015})}\BibitemShut {NoStop}%
\bibitem [{\citenamefont {Jacobson}\ and\ \citenamefont
  {Whitelam}(2019)}]{jacobson2019vard}%
  \BibitemOpen
  \bibfield  {author} {\bibinfo {author} {\bibfnamefont {D.}~\bibnamefont
  {Jacobson}}\ and\ \bibinfo {author} {\bibfnamefont {S.}~\bibnamefont
  {Whitelam}},\ }\href@noop {} {\enquote {\bibinfo {title}
  {vard\_correction},}\ }\bibinfo {howpublished}
  {\url{https://github.com/drdrjacobs/vard_correction.git}} (\bibinfo {year}
  {2019})\BibitemShut {NoStop}%
\bibitem [{\citenamefont {Torrie}\ and\ \citenamefont
  {Valleau}(1977)}]{torrie1977nonphysical}%
  \BibitemOpen
  \bibfield  {author} {\bibinfo {author} {\bibfnamefont {G.}~\bibnamefont
  {Torrie}}\ and\ \bibinfo {author} {\bibfnamefont {J.}~\bibnamefont
  {Valleau}},\ }\href@noop {} {\bibfield  {journal} {\bibinfo  {journal}
  {Journal of Computational Physics}\ }\textbf {\bibinfo {volume} {23}},\
  \bibinfo {pages} {187} (\bibinfo {year} {1977})}\BibitemShut {NoStop}%
\bibitem [{\citenamefont {ten Wolde}\ and\ \citenamefont
  {Frenkel}(1998)}]{tenWolde_1998}%
  \BibitemOpen
  \bibfield  {author} {\bibinfo {author} {\bibfnamefont {P.~R.}\ \bibnamefont
  {ten Wolde}}\ and\ \bibinfo {author} {\bibfnamefont {D.}~\bibnamefont
  {Frenkel}},\ }\href@noop {} {\bibfield  {journal} {\bibinfo  {journal} {The
  Journal of Chemical Physics}\ }\textbf {\bibinfo {volume} {109}},\ \bibinfo
  {pages} {9901} (\bibinfo {year} {1998})}\BibitemShut {NoStop}%
\bibitem [{\citenamefont {Derrida}(1998)}]{derrida1998exactly}%
  \BibitemOpen
  \bibfield  {author} {\bibinfo {author} {\bibfnamefont {B.}~\bibnamefont
  {Derrida}},\ }\href@noop {} {\bibfield  {journal} {\bibinfo  {journal}
  {Physics Reports}\ }\textbf {\bibinfo {volume} {301}},\ \bibinfo {pages} {65}
  (\bibinfo {year} {1998})}\BibitemShut {NoStop}%
\bibitem [{\citenamefont {Fredrickson}\ and\ \citenamefont
  {Andersen}(1984)}]{fredrickson1984kinetic}%
  \BibitemOpen
  \bibfield  {author} {\bibinfo {author} {\bibfnamefont {G.~H.}\ \bibnamefont
  {Fredrickson}}\ and\ \bibinfo {author} {\bibfnamefont {H.~C.}\ \bibnamefont
  {Andersen}},\ }\href@noop {} {\bibfield  {journal} {\bibinfo  {journal}
  {Physical Review Letters}\ }\textbf {\bibinfo {volume} {53}},\ \bibinfo
  {pages} {1244} (\bibinfo {year} {1984})}\BibitemShut {NoStop}%
\bibitem [{\citenamefont {Binder}(1986)}]{binder1986introduction}%
  \BibitemOpen
  \bibfield  {author} {\bibinfo {author} {\bibfnamefont {K.}~\bibnamefont
  {Binder}},\ }in\ \href@noop {} {\emph {\bibinfo {booktitle} {Monte Carlo
  Methods in Statistical Physics}}}\ (\bibinfo  {publisher} {Springer},\
  \bibinfo {year} {1986})\ pp.\ \bibinfo {pages} {1--45}\BibitemShut {NoStop}%
\bibitem [{\citenamefont {Gillespie}(1977)}]{gillespie1977exact}%
  \BibitemOpen
  \bibfield  {author} {\bibinfo {author} {\bibfnamefont {D.~T.}\ \bibnamefont
  {Gillespie}},\ }\href@noop {} {\bibfield  {journal} {\bibinfo  {journal} {The
  journal of physical chemistry}\ }\textbf {\bibinfo {volume} {81}},\ \bibinfo
  {pages} {2340} (\bibinfo {year} {1977})}\BibitemShut {NoStop}%
\bibitem [{\citenamefont {Dinwoodie}\ and\ \citenamefont
  {Zabell}(1992)}]{dinwoodie1992large}%
  \BibitemOpen
  \bibfield  {author} {\bibinfo {author} {\bibfnamefont {I.~H.}\ \bibnamefont
  {Dinwoodie}}\ and\ \bibinfo {author} {\bibfnamefont {S.~L.}\ \bibnamefont
  {Zabell}},\ }\href@noop {} {\bibfield  {journal} {\bibinfo  {journal} {The
  Annals of Probability}\ }\textbf {\bibinfo {volume} {20}},\ \bibinfo {pages}
  {1147} (\bibinfo {year} {1992})}\BibitemShut {NoStop}%
\bibitem [{\citenamefont {Ellis}(1985)}]{ellis1985large}%
  \BibitemOpen
  \bibfield  {author} {\bibinfo {author} {\bibfnamefont {R.~S.}\ \bibnamefont
  {Ellis}},\ }\href@noop {} {\emph {\bibinfo {title} {Large deviations and
  statistical mechanics}}},\ Vol.~\bibinfo {volume} {17}\ (\bibinfo
  {publisher} {Springer, New York},\ \bibinfo {year} {1985})\BibitemShut
  {NoStop}%
\bibitem [{Note3()}]{Note3}%
  \BibitemOpen
  \bibinfo {note} {In an abuse of notation we have, for brevity, changed from
  writing the joint rate function in the form $\protect \mathaccentV
  {tilde}07E{J}(q,a)$ in (\ref {int1}) to $\protect \mathaccentV
  {tilde}07E{J}(\delta q,a)$ in (\ref {j1}) and subsequently.}\BibitemShut
  {Stop}%
\bibitem [{Note4()}]{Note4}%
  \BibitemOpen
  \bibinfo {note} {Note that the reference model can be well behaved in this
  manner even when the {\protect \em original} model exhibits anomalous
  fluctuations such that $J(a)$ is not quadratic about its minimum~\cite
  {klymko2017rare}.}\BibitemShut {Stop}%
\bibitem [{\citenamefont {Duffy}\ and\ \citenamefont
  {Metcalfe}(2005)}]{duffy2005large}%
  \BibitemOpen
  \bibfield  {author} {\bibinfo {author} {\bibfnamefont {K.}~\bibnamefont
  {Duffy}}\ and\ \bibinfo {author} {\bibfnamefont {A.~P.}\ \bibnamefont
  {Metcalfe}},\ }\href@noop {} {\bibfield  {journal} {\bibinfo  {journal}
  {Journal of applied probability}\ }\textbf {\bibinfo {volume} {42}},\
  \bibinfo {pages} {267} (\bibinfo {year} {2005})}\BibitemShut {NoStop}%
\bibitem [{Note5()}]{Note5}%
  \BibitemOpen
  \bibinfo {note} {The choice of moving in the positive $k_a$ direction is
  arbitrary. Moving in the negative $k_{a}$ direction will yield an equivalent
  convergence criterion (similarly we can move in either the positive or
  negative $k_{\delta q}$ directions while holding $k_a = 0$).}\BibitemShut
  {Stop}%
\bibitem [{Note6()}]{Note6}%
  \BibitemOpen
  \bibinfo {note} {The quantity $\protect \mathaccentV {hat}05E{\protect
  \mathaccentV {tilde}07E{J}}(a')$ on the right-side of (\ref
  {convergence_check}) is determined from the Legendre transform (\ref {lt}) of
  $\protect \mathaccentV {hat}05E{\protect \mathaccentV {tilde}07E{\theta
  }}(k_{\delta q} = 0, k_a = \protect \mathaccentV {hat}05E{k}_a')$. The
  quantity $\protect \mathaccentV {hat}05E{k}_a'$ is calculated by finding the
  maximum of $\protect \mathrm {Err}[\protect \mathaccentV {hat}05E{a}(k_a)]$,
  Eq.~(\ref {error_a}), while starting at the point $(k_{\delta q} = 0, k_a =
  0)$ on the SCGF and scanning outwards by increasing $k_a$.}\BibitemShut
  {Stop}%
\bibitem [{\citenamefont {Budini}\ \emph {et~al.}(2014)\citenamefont {Budini},
  \citenamefont {Turner},\ and\ \citenamefont
  {Garrahan}}]{budini2014fluctuating}%
  \BibitemOpen
  \bibfield  {author} {\bibinfo {author} {\bibfnamefont {A.~A.}\ \bibnamefont
  {Budini}}, \bibinfo {author} {\bibfnamefont {R.~M.}\ \bibnamefont {Turner}},
  \ and\ \bibinfo {author} {\bibfnamefont {J.~P.}\ \bibnamefont {Garrahan}},\
  }\href@noop {} {\bibfield  {journal} {\bibinfo  {journal} {Journal of
  Statistical Mechanics: Theory and Experiment}\ }\textbf {\bibinfo {volume}
  {2014}},\ \bibinfo {pages} {P03012} (\bibinfo {year} {2014})}\BibitemShut
  {NoStop}%
\bibitem [{\citenamefont {Derrida}\ \emph {et~al.}(2002)\citenamefont
  {Derrida}, \citenamefont {Lebowitz},\ and\ \citenamefont
  {Speer}}]{derrida2002large}%
  \BibitemOpen
  \bibfield  {author} {\bibinfo {author} {\bibfnamefont {B.}~\bibnamefont
  {Derrida}}, \bibinfo {author} {\bibfnamefont {J.}~\bibnamefont {Lebowitz}}, \
  and\ \bibinfo {author} {\bibfnamefont {E.}~\bibnamefont {Speer}},\
  }\href@noop {} {\bibfield  {journal} {\bibinfo  {journal} {Journal of
  statistical physics}\ }\textbf {\bibinfo {volume} {107}},\ \bibinfo {pages}
  {599} (\bibinfo {year} {2002})}\BibitemShut {NoStop}%
\bibitem [{\citenamefont {Kolomeisky}\ \emph {et~al.}(1998)\citenamefont
  {Kolomeisky}, \citenamefont {Sch{\"u}tz}, \citenamefont {Kolomeisky},\ and\
  \citenamefont {Straley}}]{kolomeisky1998phase}%
  \BibitemOpen
  \bibfield  {author} {\bibinfo {author} {\bibfnamefont {A.~B.}\ \bibnamefont
  {Kolomeisky}}, \bibinfo {author} {\bibfnamefont {G.~M.}\ \bibnamefont
  {Sch{\"u}tz}}, \bibinfo {author} {\bibfnamefont {E.~B.}\ \bibnamefont
  {Kolomeisky}}, \ and\ \bibinfo {author} {\bibfnamefont {J.~P.}\ \bibnamefont
  {Straley}},\ }\href@noop {} {\bibfield  {journal} {\bibinfo  {journal}
  {Journal of Physics A: Mathematical and General}\ }\textbf {\bibinfo {volume}
  {31}},\ \bibinfo {pages} {6911} (\bibinfo {year} {1998})}\BibitemShut
  {NoStop}%
\bibitem [{\citenamefont {Blythe}\ \emph {et~al.}(2000)\citenamefont {Blythe},
  \citenamefont {Evans}, \citenamefont {Colaiori},\ and\ \citenamefont
  {Essler}}]{blythe2000exact}%
  \BibitemOpen
  \bibfield  {author} {\bibinfo {author} {\bibfnamefont {R.}~\bibnamefont
  {Blythe}}, \bibinfo {author} {\bibfnamefont {M.}~\bibnamefont {Evans}},
  \bibinfo {author} {\bibfnamefont {F.}~\bibnamefont {Colaiori}}, \ and\
  \bibinfo {author} {\bibfnamefont {F.}~\bibnamefont {Essler}},\ }\href@noop {}
  {\bibfield  {journal} {\bibinfo  {journal} {Journal of Physics A:
  Mathematical and General}\ }\textbf {\bibinfo {volume} {33}},\ \bibinfo
  {pages} {2313} (\bibinfo {year} {2000})}\BibitemShut {NoStop}%
\bibitem [{\citenamefont {Garrahan}\ and\ \citenamefont
  {Chandler}(2002)}]{garrahan2002geometrical}%
  \BibitemOpen
  \bibfield  {author} {\bibinfo {author} {\bibfnamefont {J.~P.}\ \bibnamefont
  {Garrahan}}\ and\ \bibinfo {author} {\bibfnamefont {D.}~\bibnamefont
  {Chandler}},\ }\href@noop {} {\bibfield  {journal} {\bibinfo  {journal}
  {Physical Review Letters}\ }\textbf {\bibinfo {volume} {89}},\ \bibinfo
  {pages} {035704} (\bibinfo {year} {2002})}\BibitemShut {NoStop}%
\bibitem [{\citenamefont {Ba{\~n}uls}\ and\ \citenamefont
  {Garrahan}(2019)}]{banuls2019using}%
  \BibitemOpen
  \bibfield  {author} {\bibinfo {author} {\bibfnamefont {M.~C.}\ \bibnamefont
  {Ba{\~n}uls}}\ and\ \bibinfo {author} {\bibfnamefont {J.~P.}\ \bibnamefont
  {Garrahan}},\ }\href@noop {} {\bibfield  {journal} {\bibinfo  {journal}
  {arXiv preprint arXiv:1903.01570}\ } (\bibinfo {year} {2019})}\BibitemShut
  {NoStop}%
\bibitem [{\citenamefont {Whitelam}\ \emph {et~al.}(2019)\citenamefont
  {Whitelam}, \citenamefont {Jacobson},\ and\ \citenamefont
  {Tamblyn}}]{whitelam2019evolutionary}%
  \BibitemOpen
  \bibfield  {author} {\bibinfo {author} {\bibfnamefont {S.}~\bibnamefont
  {Whitelam}}, \bibinfo {author} {\bibfnamefont {D.}~\bibnamefont {Jacobson}},
  \ and\ \bibinfo {author} {\bibfnamefont {I.}~\bibnamefont {Tamblyn}},\
  }\href@noop {} {\bibfield  {journal} {\bibinfo  {journal} {arXiv preprint
  arXiv:1909.00835}\ } (\bibinfo {year} {2019})}\BibitemShut {NoStop}%
\bibitem [{Note7()}]{Note7}%
  \BibitemOpen
  \bibinfo {note} {The method works for Markov chains in discrete time, upon
  replacing the distributions (\ref {two}) and (\ref {four}) by $\delta (\Delta
  t-1)$; see earlier versions of the method~\cite
  {klymko2017rare,whitelam2018sampling,whitelam2018multi}.}\BibitemShut {Stop}%
\bibitem [{\citenamefont {Frenkel}\ and\ \citenamefont
  {Ladd}(1984)}]{frenkel1984new}%
  \BibitemOpen
  \bibfield  {author} {\bibinfo {author} {\bibfnamefont {D.}~\bibnamefont
  {Frenkel}}\ and\ \bibinfo {author} {\bibfnamefont {A.~J.}\ \bibnamefont
  {Ladd}},\ }\href@noop {} {\bibfield  {journal} {\bibinfo  {journal} {The
  Journal of Chemical Physics}\ }\textbf {\bibinfo {volume} {81}},\ \bibinfo
  {pages} {3188} (\bibinfo {year} {1984})}\BibitemShut {NoStop}%
\bibitem [{Note8()}]{Note8}%
  \BibitemOpen
  \bibinfo {note} {One advantage of bypassing the SCGF and calculating the rate
  function directly, as we do here, is the ability to reconstruct rate
  functions that are not strictly convex~\cite
  {touchette2009large,klymko2017rare,nickelsen2018anomalous}.}\BibitemShut
  {Stop}%
\bibitem [{\citenamefont {Spiliopoulos}(2013)}]{spiliopoulos2013large}%
  \BibitemOpen
  \bibfield  {author} {\bibinfo {author} {\bibfnamefont {K.}~\bibnamefont
  {Spiliopoulos}},\ }\href@noop {} {\bibfield  {journal} {\bibinfo  {journal}
  {Applied Mathematics \& Optimization}\ }\textbf {\bibinfo {volume} {67}},\
  \bibinfo {pages} {123} (\bibinfo {year} {2013})}\BibitemShut {NoStop}%
\bibitem [{\citenamefont {Bouchet}\ \emph {et~al.}(2016)\citenamefont
  {Bouchet}, \citenamefont {Grafke}, \citenamefont {Tangarife},\ and\
  \citenamefont {Vanden-Eijnden}}]{bouchet2016large}%
  \BibitemOpen
  \bibfield  {author} {\bibinfo {author} {\bibfnamefont {F.}~\bibnamefont
  {Bouchet}}, \bibinfo {author} {\bibfnamefont {T.}~\bibnamefont {Grafke}},
  \bibinfo {author} {\bibfnamefont {T.}~\bibnamefont {Tangarife}}, \ and\
  \bibinfo {author} {\bibfnamefont {E.}~\bibnamefont {Vanden-Eijnden}},\
  }\href@noop {} {\bibfield  {journal} {\bibinfo  {journal} {Journal of
  Statistical Physics}\ }\textbf {\bibinfo {volume} {162}},\ \bibinfo {pages}
  {793} (\bibinfo {year} {2016})}\BibitemShut {NoStop}%
\bibitem [{\citenamefont {Whitelam}(2018{\natexlab{c}})}]{whitelam2018large}%
  \BibitemOpen
  \bibfield  {author} {\bibinfo {author} {\bibfnamefont {S.}~\bibnamefont
  {Whitelam}},\ }\href@noop {} {\bibfield  {journal} {\bibinfo  {journal}
  {Physical Review E}\ }\textbf {\bibinfo {volume} {97}},\ \bibinfo {pages}
  {062109} (\bibinfo {year} {2018}{\natexlab{c}})}\BibitemShut {NoStop}%
\bibitem [{\citenamefont {Nickelsen}\ and\ \citenamefont
  {Touchette}(2018)}]{nickelsen2018anomalous}%
  \BibitemOpen
  \bibfield  {author} {\bibinfo {author} {\bibfnamefont {D.}~\bibnamefont
  {Nickelsen}}\ and\ \bibinfo {author} {\bibfnamefont {H.}~\bibnamefont
  {Touchette}},\ }\href@noop {} {\bibfield  {journal} {\bibinfo  {journal}
  {arXiv preprint arXiv:1803.05708}\ } (\bibinfo {year} {2018})}\BibitemShut
  {NoStop}%
\end{thebibliography}

%

\end{document}